%% file: crosscalib_revised.tex
\documentclass[traditabstract, longauth]{aa} 

\usepackage[english]{babel} 
\usepackage[utf8]{inputenc} 
\usepackage[T1]{fontenc}

\usepackage{natbib}
\usepackage{amsmath}
\usepackage{txfonts}
\usepackage{graphicx}
\usepackage{epstopdf}
\usepackage[breaklinks, colorlinks,citecolor=blue]{hyperref} 
\usepackage[usenames,dvipsnames,table]{xcolor}
\usepackage{longtable,lscape}
\usepackage{units}
\bibpunct{(}{)}{;}{a}{}{,}  

\input Planck.tex

\graphicspath{{../figures/}{./figures/}}

\def\bfc{}

\def\reff@jnl#1{{#1\/}}
\def\apj{\reff@jnl{ApJ}}       
\def\apjs{\reff@jnl{ApJS}}     
\def\aaps{\reff@jnl{A\&AS}}    
\def\mnras{\reff@jnl{MNRAS}}   
\def\prd{\reff@jnl{Phys.\ Rev.\ D}}    

{\begin{enumerate}\setlength{\itemsep}{0mm}}%
{\end{enumerate}}
{\begin{enumerate}\setlength{\itemsep}{0mm}}%
{\end{enumerate}}

\newcommand{\bc}{\begin{center}}
\newcommand{\ec}{\end{center}}
\newcommand{\bi}{\begin{itemize}}
\newcommand{\ei}{\end{itemize}}
\newcommand{\ben}{\begin{enumerate}}
\newcommand{\een}{\end{enumerate}}

\newfont{\gwpfont}{cmssq8 scaled 1000}

\def\Jybeam{Jy beam$^{-1}$}
\def\PtoEunits{\MJysr\ Jy$^{-1}$}

\def\lfn{L59} 
\def\lth{L30} 

\def\rsf{\mathcal{N}} 

\def\GHz{\ifmmode $GHz$\else \,GHz\fi}
\def\microm{\ifmmode \,\mu$m$\else \,$\mu$\hbox{m}\fi}

\def\Herschel{\textit{Herschel}}

\newcommand{\gcinq}{1.047}
\newcommand{\egcinq}{($\pm\, 0.0069$)}
\newcommand{\ghuit}{1.003}
\newcommand{\eghuit}{($\pm\, 0.0080$)}

\begin{document}

\title{Comparison of absolute gain photometric calibration between \textit{Planck}/HFI and \Herschel/SPIRE at 545 and 857\GHz}
\author{B.~Bertincourt\inst{1} \and
	    G.~Lagache\inst{2,1} \and 
	    P.\,G.~Martin\inst{3} \and
	    B.~Schulz\inst{4} \and
	    L.~Conversi \inst{5} \and
	    K.~Dassas\inst{1} \and
	    L.~Maurin\inst{6} \and
	    A.~Abergel\inst{1} \and
	    A.~Beelen\inst{1} \and
	    J-P.~Bernard\inst{7} \and
	    B.\,P.~Crill \inst{8} \and
	    H.~Dole\inst{1} \and
	    S.~Eales\inst{9} \and
	    J.\,E.~Gudmundsson\inst{10} \and
	    E. ~Lellouch\inst{11} \and
	    R. ~Moreno\inst{11} \and
	    O.~Perdereau\inst{12}}
	     
\institute{
Institut d'Astrophysique Spatiale (IAS), B\^atiment 121, F-91405 Orsay
(France); Universit\'e Paris-Sud 11 and CNRS (UMR 8617) \goodbreak \and
Aix Marseille Universit\'e, CNRS, LAM (Laboratoire d'Astrophysique de
Marseille) UMR 7326, 13388, Marseille, France, \email{guilaine.lagache@lam.fr}\goodbreak \and
Canadian Institute for Theoretical Astrophysics, University of
Toronto, 60 St. George Street, Toronto, ON M5S 3H8, Canada \goodbreak\and
NASA \Herschel\ Science Center, IPAC, 770 South Wilson Avenue,
Pasadena, CA 91125, USA \goodbreak \and
European Space Astronomy Centre (ESAC)/ESA, Villanueva de la Canada,
E-28691 Madrid, Spain \goodbreak \and
APC, Universit\'e Paris 7 Denis Diderot, 10, rue Alice Domon et
L\'eonie Duquet, 75205 Paris Cedex 13, France \goodbreak \and
CNRS, IRAP, 9 Av. Colonel Roche, BP 44346, F-31028 Toulouse Cedex 4,
France; Universite de Toulouse, UPS-OMP, IRAP, F-31028 Toulouse Cedex
4, France \goodbreak \and
Department of Physics, California Institute of Technology, Pasadena,
CA, U.S.A.; Jet Propulsion Laboratory, California Institute of
Technology, 4800 Oak Grove Drive, Pasadena, CA, U.S.A.. \goodbreak
\and
School of Physics and Astronomy, Cardiff University, Queens Buildings,
The Parade, Cardiff CF24 3AA, UK \goodbreak \and
Department of Physics, Princeton University, Princeton, NJ,
U.S.A. \goodbreak \and
LESIA, Observatoire de Paris, CNRS, UPMC, Universit\'{e}
Paris-Diderot, 5 Place J. Janssen, 92195 Meudon, France \goodbreak
\and
Laboratoire de l'Acc\'el\'erateur Lin\'eaire, Universit\'e Paris-Sud,
CNRS/IN2P3, 91898, Orsay, France \goodbreak }
	     
\date{}

\abstract{ 
We compare the absolute gain photometric calibration of the
\textit{Planck}/HFI and \Herschel/SPIRE instruments on diffuse emission.
The absolute calibration of HFI and SPIRE each relies on planet
flux measurements and comparison with theoretical far-infrared
emission models of planetary atmospheres.
We measure the photometric cross calibration between the instruments
at two overlapping bands, 545\GHz\ / 500\microm\ and 857\GHz\ /
350\microm.
The SPIRE maps used have been processed in the \Herschel\ Interactive
Processing Environment (Version 12) and the HFI data are from the 2015
Public Data Release 2.
For our study we used 15 large fields observed with SPIRE, which cover
a total of about 120 deg$^2$. We have selected these fields carefully
to provide high signal-to-noise ratio, avoid residual systematics in
the SPIRE maps, and span a wide range of surface brightness.
The HFI maps are bandpass-corrected to match the emission observed by
the SPIRE bandpasses. The SPIRE maps are  convolved to match the HFI beam
and put on a common pixel grid. 
We measure the cross-calibration relative gain between the instruments using two methods
in each field, pixel-to-pixel correlation and angular power spectrum
measurements. 
The SPIRE\,/\,HFI relative gains are
\gcinq\,\egcinq\ and \ghuit\,\eghuit\ at 545 and 857\GHz,
respectively, indicating very good agreement between the instruments.
These relative gains deviate from unity by much less
than the uncertainty of the absolute extended emission
calibration, which is about 6.4\,\% and 9.5\,\% for HFI and SPIRE,
respectively, but the deviations are comparable to the values 1.4\,\%
and 5.5\,\% for HFI and SPIRE if the uncertainty from models of the
common calibrator can be discounted.
Of the 5.5\,\% uncertainty for SPIRE, 4\,\% arises from the uncertainty of the effective beam solid angle, which impacts the adopted
SPIRE point source to extended source unit conversion factor,
highlighting that as a focus for refinement.
}


\authorrunning{Bertincourt et al.}
\titlerunning{\textit{Planck}/HFI and \Herschel/SPIRE Cross Calibration}

\maketitle

\section{Introduction}
\label{sect:intro}

The \textit{Planck}\footnote{
\textit{Planck} (\url{http://www.esa.int/Planck}) is a project of the
European Space Agency (ESA) with instruments provided by two
scientific consortia funded by ESA member states and led by Principal
Investigators from France and Italy, telescope reflectors provided
through a collaboration between ESA and a scientific consortium led
and funded by Denmark, and additional contributions from NASA (USA).
} 
and \Herschel\footnote{
\Herschel\ (\url{http://www.esa.int/Herschel}) is an ESA space
observatory with science instruments provided by European-led
Principal Investigator consortia and with important participation from
NASA.
}
satellites have an interconnected history, including their launch
together in 2009.
Although operating with different scientific goals across many
frequencies, \textit{Planck} and \Herschel\ have in common two very similar
passbands, 545\GHz\ / 500\microm\ and 857\GHz\ / 350\microm.  This
redundancy is very important for complementary studies.  For
example, the combination of \textit{Planck} large-scale and \Herschel\
small-scale observations is valuable to the study of the cosmic
infrared background anisotropies \citep[e.g.,][]{planck2011-6.6,
  Viero:2013fk} and to understanding the interplay between interstellar medium 
structures and their environment \citep[e.g.,][]{planck2011-7.7a,
  Juvela:2011uq}.
With the 2015 Public Data Release 2 (PR2) of the \textit{Planck} data and the
growing public availability of processed \Herschel\ observations, this
is an opportune time to address the important question of the
compatibility of measurements carried out by both instruments.

The \textit{Planck} High Frequency Instrument (HFI, \citealp{planck2013-p03})
is composed of a set of 52 bolometers observing the sky at six
frequencies (100, 143, 217, 353, 545, and 857\GHz). It provided
full-sky maps with an angular resolution ranging from $9\parcm7$ to
$4\parcm6$.  The Spectral and Photometric Imaging Receiver (SPIRE,
\citealp{Griffin:2010ys})\footnote{
The SPIRE Handbook is available at
\url{http://herschel.esac.esa.int/Docs/SPIRE/spire_handbook.pdf}.}
provided photometric capabilities at 250, 350, and 500\microm\ (PSW,
PMW, and PLW bands, respectively) to the \Herschel\ Space Observatory
\citep{Pilbratt:2010fk}.  The three arrays contain 139, 88, and 43
detectors with angular resolution of 18\parcs2, 24\parcs9, and
36\parcs3, respectively, and a common field of view of $4\arcm \times
8\arcm$.  Larger fields were scanned and maps made from the
time-stream data using software available in Version 12 of the \Herschel\
Interactive Processing Environment ({\tt
HIPE})\footnote{\url{http://herschel.esac.esa.int/hipe/}}.

In this paper we compare the absolute photometric calibration of the
HFI and SPIRE instruments in the overlapping bandpasses
(Fig.\,\ref{fig:rsrfs}), focusing on the relative
calibration \bfc{(the cross-calibration relative gain)}
for emission that is extended on the sky, i.e., the so-called diffuse
emission. The relative offset calibration will be presented
separately (Schulz et al., in preparation)\footnote{
For a preview, see
\url{http://herschel.esac.esa.int/TheUniverseExploredByHerschel/posters/B23_SchulzB.pdf}.}.

The paper is organized as follows. 
In Sect.\,\ref{sect:calib} and Appendix\,\ref{app:calib} we briefly summarize the calibration
schemes of \textit{Planck}/HFI and \Herschel/SPIRE frequency maps with
relevant details concerning beams, bandpasses, and absolute
calibrators.
In Sect.\,\ref{sect:fselect} we describe the fields that we selected
to study the cross calibration of the diffuse emission.
In Sect.\,\ref{sect:ccolor} we discuss the factors that have to be
taken into account to compare the intensity of a given source of
emission in the different bandpasses of the two instruments.
We then carry out two independent assessments of the relative gain using
maps of diffuse emission: a direct correlation analysis in
Sect.\,\ref{sect:pixtopix} and a power spectrum comparison in
Sect.\,\ref{sect:powspec}.
We conclude in Sect.\,\ref{conclu}.

\section{Absolute photometric calibration}
\label{sect:calib}
  
Following the
tradition of infrared and submillimetre experiments, pipeline
processed \textit{Planck} and \Herschel\ measurements are calculated,
calibrated, and reported as surface brightnesses (or monochromatic
flux densities)
denoted $\tilde{S}_{\nu_0}$\footnote{
The spectral energy distribution (the SED) is often denoted $I_\nu$ in the case of
surface brightness and $F_\nu$ in
the case of point source flux densities.  In this paper we will use $S_\nu$ for both, the
distinction being clear from the context. 
\bfc{The pipeline product that we denote $\tilde{S}_{\nu_0}$ is called $S_{\mathrm{pip}}$ in the SPIRE Handbook.}
},
adopting as a reference
spectrum a power law of index $-1$ (i.e., $\nu\ S_{\text{ref}}(\nu) =
\text{constant}$).  Thus
\begin{equation}
\tilde{S}_{\nu_0} = \frac{1}{\nu_0} \frac{\int \rsf(\nu)\,S(\nu) \ d\nu}{\int \rsf(\nu)/\nu \ d\nu} \,,
\label{eq:snu0}
\end{equation}
where $\nu_0$ is the adopted nominal frequency of the filter.
For the standard HFI and SPIRE pipelines, $\nu_0$ is chosen to be equal to 545 and 857\,\GHz, and to correspond to
wavelengths of 350  and 500\microm, respectively. The parameter $\rsf(\nu)$ is the 
\bfc{net spectral response in terms of absorbed power
including the aperture efficiency, the filter spectral response, and for surface brightness the spectral dependence of the beam across the band (see, e.g., details in \citealp{Griffin:2013vn} and the SPIRE Handbook, Chapter~5). Its normalization is unimportant because the response is used only in ratios.} 
Absolute calibration therefore requires precise knowledge of the instrument properties, particularly
bandpasses and beams, and there are related uncertainties summarized here.   See Appendix\,\ref{app:calib} for more details.

\begin{figure}
\begin{center}
\includegraphics[width=7.8cm]{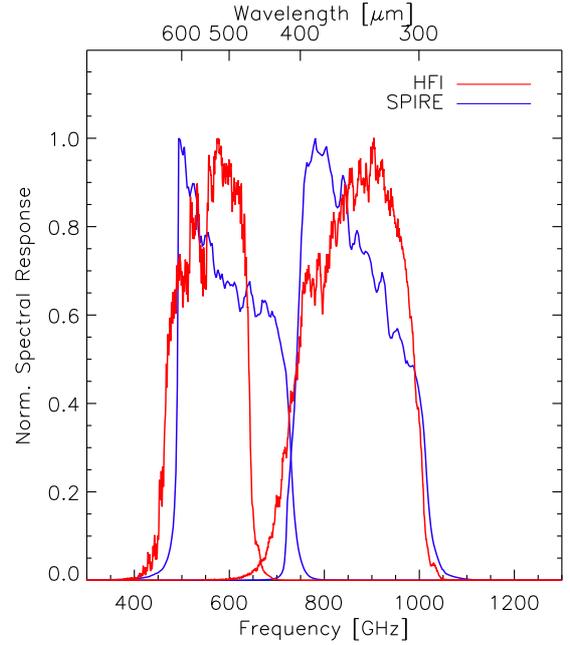}
\caption{Spectral response for the \textit{Planck}/HFI 545 and 857\GHz\ filters (red) and the \Herschel/SPIRE PLW
and PMW filters (blue), respectively, \bfc{each normalized relative to a maximum at unity.}  For SPIRE these 
\bfc{include the spectral dependence of the beam across the band as is appropriate for extended emission (compare Fig.\,5.16 to Fig.\,5.5 in the SPIRE Handbook v2.5).}
No correction is needed for HFI.
}
\label{fig:rsrfs}
\end{center}
\end{figure}

Recovering the true monochromatic brightness (or flux density),
$S_{\nu_0}$, from $\tilde{S}_{\nu_0}$ requires an instrument-dependent colour correction
that depends on the shape of the SED, which varies over the sky.  
Because we are only interested in the relative calibration of the two instruments, 
these colour corrections are not needed explicitly. However, because of the different shapes of the
spectral response functions for the pairs of bandpasses of the two instruments we are comparing
(Fig.\,\ref{fig:rsrfs}) and the different values of $\nu_0$, a
bandpass correction dependent on the shape of the SED is needed for
the cross-calibration analysis (see Sect.\,\ref{sect:ccolor}).

The \textit{Planck}/HFI calibration scheme for the 2015 PR2 data is described
in \cite{planck2014-a09}. 
The 545 and 857\GHz\ channels were calibrated by
comparing measurements of Uranus and Neptune flux densities 
with predictions based on models of their emissivities
(the so-called Uranus ESA\,2 and Neptune ESA\,3 models produced by
\citealp{Moreno2010} updating \citealp{Moreno:1998PhD}).
The statistical uncertainty of the measurements is 1.1\,\% and 1.4\,\%
at 545 and 857\GHz, respectively, and 
the model predictions have an absolute uncertainty of 5\,\%.
Combining the statistical and systematic uncertainties linearly, the overall
HFI absolute calibration uncertainty for PR2 data is 6.1\,\% and
6.4\,\% at 545 and 857\GHz, respectively.

The relative planet model uncertainty between these two HFI bands,
however, is expected to be of order 2\,\%.  Combining the statistical
uncertainties in quadrature (thus 1.8\,\%) with this relative planet model uncertainty linearly, the
relative uncertainty between the two bands would be about 3.8\,\%.

As small bolometer to bolometer differences and variations over the sky have a negligible effect on our results (Appendix\,\ref{sect:hficalib}),
we used a Gaussian beam with a full width half maximum (FWHM) of 4\parcm83 and 4\parcm64 at 545 and
857\GHz, respectively \citep{planck2014-a08}.

The spectral responses shown in Fig.\,\ref{fig:rsrfs} are part of the \textit{Planck} data
release. We do not consider their uncertainties in this
study; reported uncertainties on planet colour correction
factors (Mars, Jupiter, Saturn, Uranus, Neptune) derived from these
spectral responses are about $0.01\,\%$ and $0.014\,\%$ at 545 and 857\GHz,
respectively \citep{planck2013-p03d}.

\citet{planck2014-a10} found that, with respect to the very accurate planet-independent calibration of the 100 and 143~GHz
channels, the relative photometric
calibration of the 545~GHz channel is  $+2.3$\,\%$\,\pm\,$1.6\,\% using the solar 
dipole and $+1.5$\,\%$\,\pm\,$1.8\,\% using the first two CMB acoustic peaks (Appendix\,\ref{sect:hficalib}).
This suggests that the uncertainty of the planet-based absolute
calibration of the 545~GHz channel is not as great as the value of 6.1\,\%
cited above, of which 5\,\% arose from the absolute uncertainty of the planet model.

The photometric calibration of the \Herschel/SPIRE instrument is described
in detail in \cite{Bendo:2013fk} and in the SPIRE Handbook.
Calibration on the ESA\,4 model of Neptune
\citep{Moreno2012} introduces a 4\,\% systematic uncertainty from
model predictions. The statistical
uncertainty on SPIRE photometry of the calibrators is about 1.5\,\%
\citep[cf.][]{Bendo:2013fk} in the 350\microm\ (PMW) and 500\microm\
(PLW) bands.  These two contributions add linearly to produce a
5.5\,\% uncertainty in the point source calibration.

The pipeline processing used to produce the SPIRE maps needed in this paper made use of a conversion from point source flux density [\Jybeam] to
extended source surface brightness [\MJysr] \citep{Griffin:2013vn}, again for a reference
spectrum $\nu\ S_{\text{ref}}(\nu) = \text{constant}$.
In the notation used in the SPIRE Handbook v2.5 the conversion factor may be written as 
$K_{\text{PtoE}} = (K_{\text{4E}}/K_{\text{4P}})/\Omega_{\text{eff}}$
to highlight the inverse dependence on $\Omega_{\text{eff}}$, which is the so-called effective beam solid angle\footnote{This solid angle results from integration of the beam over $4\pi$ steradians, as distinct from the main beam solid angle for the main lobe, and hence corresponds to what elsewhere is called the antenna beam solid angle (see, e.g., \url{http://ipnpr.jpl.nasa.gov/progress_report/42-64/64T.PDF}).}
for that bandpass accounting for the frequency dependence of the beam and calculated for the reference spectrum.
The ratio $K_{\text{4E}}/K_{\text{4P}}$ of monochromatic conversion factors defined in the Handbook, also calculated for this reference spectrum, is close to unity.
Following Table~5.2 of the SPIRE Handbook v2.5 as incorporated in the {\tt HIPE} Version 12, 
the values of this ratio and $\Omega_{\text{eff}}$
that we used are $1.0015$ and $0.9993$  and
$822.58\,\textrm{arcsec}^2$ and $1768.66\,\textrm{arcsec}^2$, so that
$K_{\text{PtoE}} $ is $51.799$ and
$24.039$\, \PtoEunits\ for PMW and PLW, respectively.   
Underlying spectral responses of SPIRE for extended emission are shown in
Fig.\,\ref{fig:rsrfs}.

The uncertainty of $K_{\text{PtoE}}$ is dominated by the 4\,\%
uncertainty of the effective beam solid angles \citep{Griffin:2013vn}. Adding this to
the 5.5\,\% uncertainty of the point source calibration, the total
uncertainty of the extended source brightness calibration amounts to
9.5\,\%.
Ongoing efforts to improve the modelling of the radial beam profile are described in Appendix\,\ref{sect:spicalib} and the implication of any change is quantified in Sect.\,\ref{conclu}.

Although Neptune is the reference calibrator in common with both
instruments (Uranus is also used for HFI), two different versions of
the ESA model have been used. 
As described in Appendix\,\ref{sect:neptmod}, the net effect of using the ESA4 model for the HFI calibration, combined with the absolute photometric
calibration derived from Uranus model, would be an increase of HFI brightness
of 0.31\,\% and 0.16\,\% at 545 and 857\GHz, respectively.  Compared to the estimated uncertainties in 
cross-calibration relative gains found below in
Sect.\,\ref{ssect:gains}, this systematic uncertainty is not a major concern.

\begin{table*}
\begingroup 
\newdimen\tblskip \tblskip=5pt
\caption{SPIRE fields selected for the SPIRE--HFI cross-calibration study.} 
\label{tab:fields}
\vskip -6mm
\footnotesize 
\setbox\tablebox=\vbox{
\newdimen\digitwidth
\setbox0=\hbox{\rm 0}
\digitwidth=\wd0
\catcode`*=\active
\def*{\kern\digitwidth}
\newdimen\signwidth
\setbox0=\hbox{+}
\signwidth=\wd0
\catcode`!=\active
\def!{\kern\signwidth}
\newdimen\decimalwidth
\setbox0=\hbox{.}
\decimalwidth=\wd0
\catcode`@=\active
\def@{\kern\signwidth}
\halign{ \hbox to 1.3in{#\leaderfil}\tabskip=1em& 
    \hfil#\hfil\tabskip=2em& 
    \hfil#\hfil& 
    \hfil#\hfil& 
    \hfil#\hfil \tabskip=0pt\cr
\noalign{\doubleline}
\omit Name & Centre (RA, Dec) & Size & $\bar{S}_\nu\text{(545\GHz)}$ &  $\bar{S}_\nu\text{(857\GHz)}$ \cr
\omit  & [degrees] & [degrees]& [\MJysr] & [\MJysr] \cr
\noalign{\vskip 3pt\hrule\vskip 5pt}
Hi-GAL \lfn            & (295.437, $*23.0302$) & 2.5 x 2.5 & *28.31 & *86.86 \cr
Hi-GAL \lth            & (281.541, $*-2.6095$) & 2.5 x 2.5 & *63.53 & 219.03 \cr
Hi-GAL Field 0\_0 & (266.422, $-28.9363$) & 2.5 x 2.5 & *73.56 & 264.37 \cr
Hi-GAL Field 2\_0 & (267.710, $-27.0518$) & 2.5 x 2.5 & *75.65 & 284.66 \cr
Hi-GAL Field 4\_0 & (268.921, $-25.1574$) & 2.5 x 2.5 & *74.53 & 270.44 \cr
Hi-GAL Field 6\_0 & (270.130, $-23.2524$) & 2.5 x 2.5 & *66.18 & 237.39 \cr
Hi-GAL Field 8\_0 & (271.302, $-21.3456$) & 2.5 x 2.5 & *66.49 & 236.40 \cr
Hi-GAL Field 33\_0 & (282.893, $**0.0483$) & 2.5 x 2.5 &  *95.67 & 323.82 \cr
Hi-GAL Field 35\_0 & (283.896, $**2.0066$) & 2.5 x 2.5 & 109.95 & 360.11 \cr
Hi-GAL Field 39\_0 & (285.909, $**5.9450$) & 2.5 x 2.5 &  *95.41 & 306.30 \cr
Hi-GAL Field 332\_0 & (243.747, $-50.9331$) & 2.5 x 2.5 & 144.86 & 508.12 \cr
Hi-GAL Field 334\_0 & (246.174, $-49.3698$) & 2.5 x 2.5 & 130.04 & 452.62 \cr
Aquila                       & (277.427, $*-2.7803$) & 3.7 x 3.9 & *29.25 & *87.14 \cr
Polaris                      & (*51.862, $*88.5574$) & 3.5 x 3.5 & **2.13 & **6.14 \cr
Spider                       & (159.490, $*72.9949$) & 4.5 x 4.2 & **0.65 & **1.81 \cr
\noalign{\vskip 5pt\hrule\vskip 3pt}
}}
\endPlancktable 
\endgroup
\end{table*} 


\begin{figure*}
\begin{center}
\begin{tabular}{cc}
\includegraphics[width=9cm]{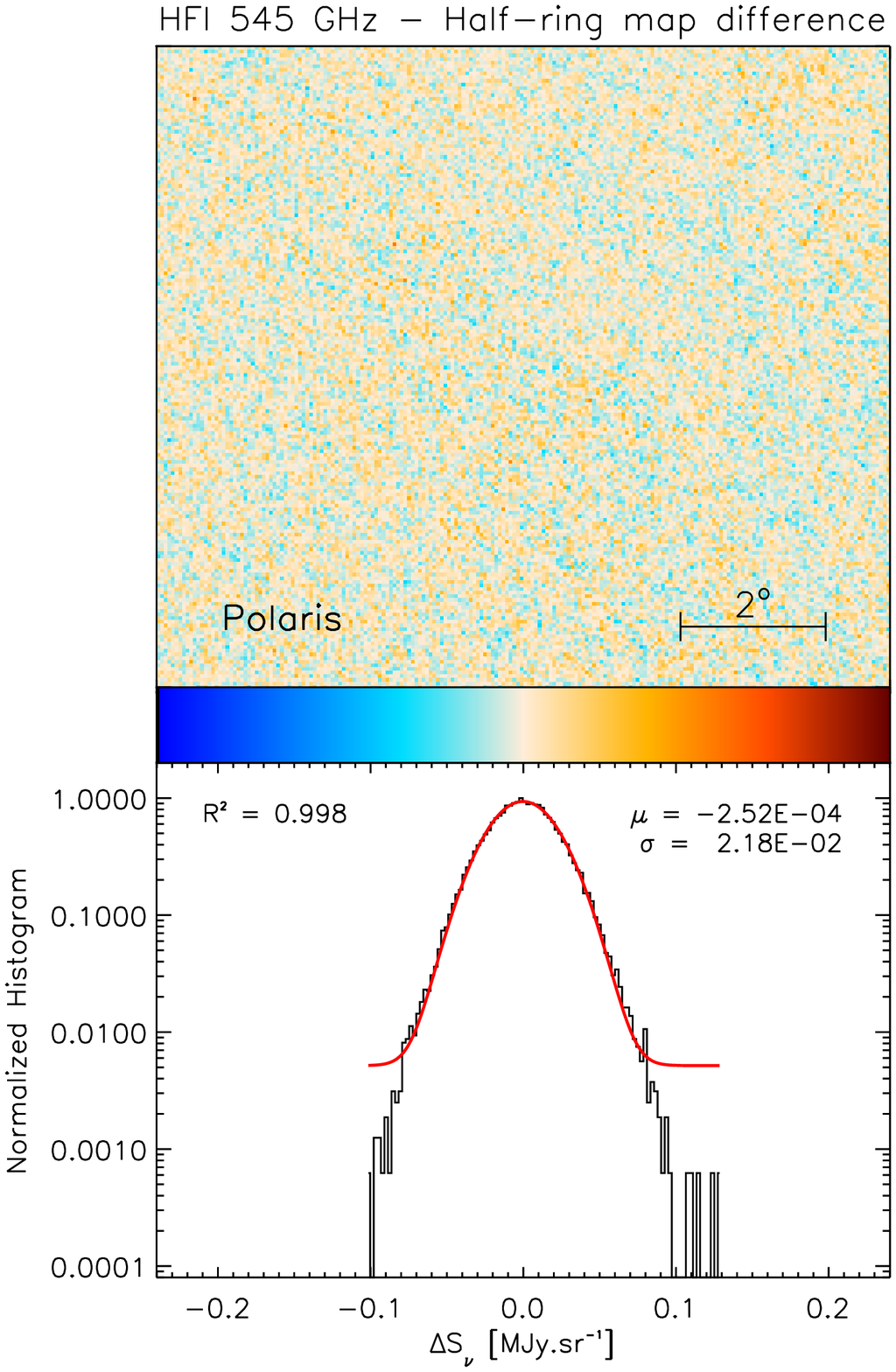} & 
\includegraphics[width=9cm]{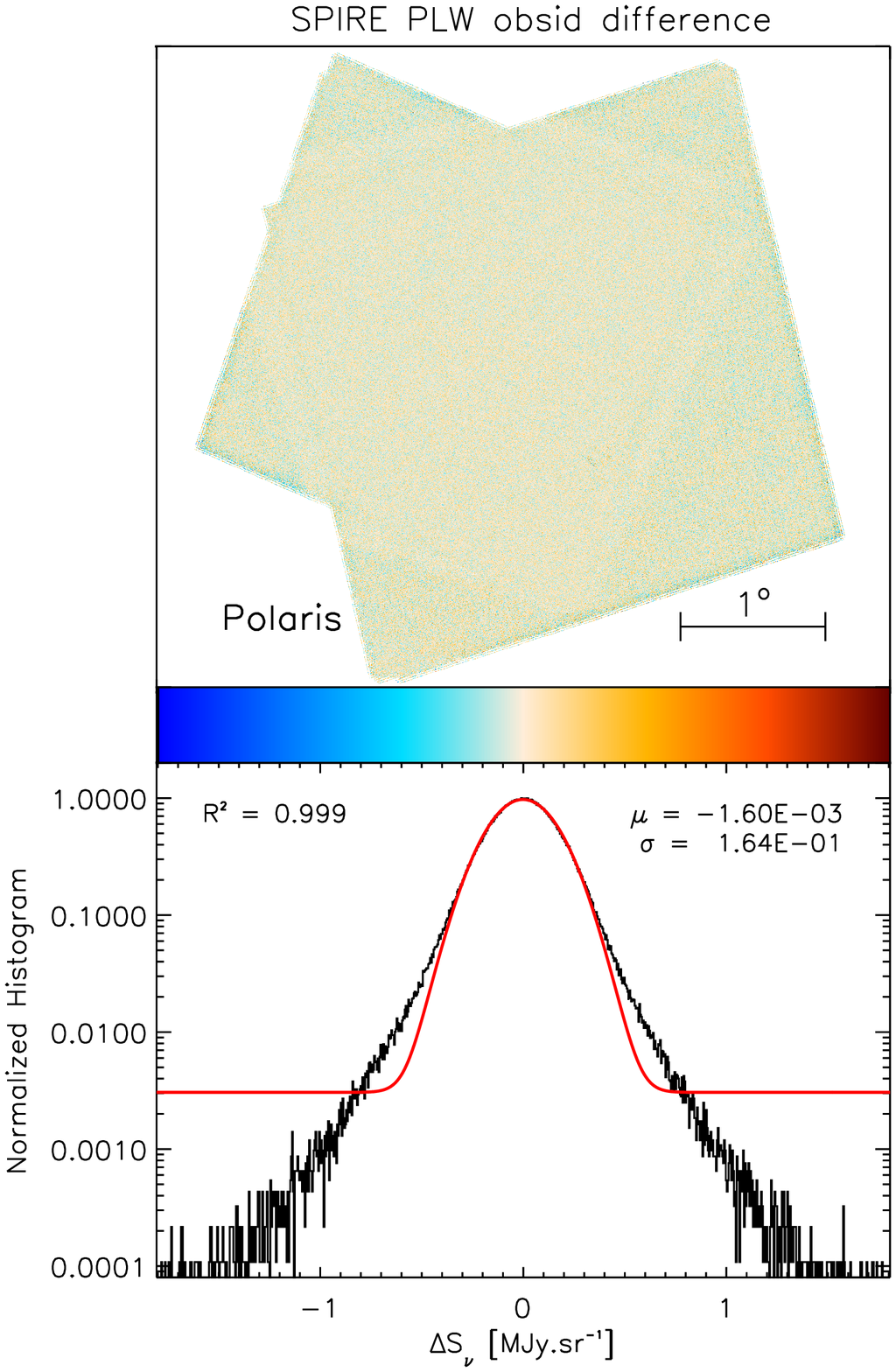} \\
\end{tabular}
\caption[caption]{
{\it Left:} HFI 545\GHz\ noise map of the Polaris field obtained from
the difference of the two independent half-ring maps of the same field (see text). 
The bottom
panel shows the normalized histogram of brightnesses in the difference
map (black) and a Gaussian fit to the distribution (red), with the
mean $\mu$, dispersion $\sigma$, and \bfc{coefficient of determination $R^2$ as an estimator of the ``goodness of fit''} \protect\footnotemark . 
{\it Right:} SPIRE PLW (500\microm) Polaris field noise map and
histogram obtained from the difference between maps produced from
scans in the nominal and orthogonal orientation, appropriately
reweighted by the coverage map.
The individual $3\deg \times 3\deg$ and $3\pdeg5 \times 3\pdeg5$
subfields were mapped at different times and
so the effective noise is lower where they overlap.  Likewise, the
largest noise occurs at the edges of the subfields where the coverage
by the bolometer arrays is least.
}
\label{fig:diff_maps}
\end{center}
\end{figure*}

\section{Field selection and map preparation}
\label{sect:fselect}

\footnotetext{The coefficient of determination provides a measure of the fraction of variance of the data explained by the model.}

From the \Herschel\ Science Archive
(HSA)\footnote{
\url{http://archives.esac.esa.int/hsa/aio/doc/}} 
we selected 15 fields covering an appropriate range of average surface
brightness and dynamics: two Science Demonstration Phase Hi-GAL fields
at $b=0\deg$ (\lfn\ centred at $l=59\deg$ and \lth\ at $l=30\deg$),
ten other fields in the Galactic plane from the Hi-GAL survey
\citep{Molinari:2010kx}, and three fields at higher latitude; these are Aquila
\citep{Konyves:2010fk}, Polaris \citep{Miville-Deschenes:2010uq}, and
Spider (in the North Celestial Pole Loop;
\citealp[e.g.,][]{Martin2015}).
Table\,\ref{tab:fields} lists the field name, centre coordinates, size,
and average surface brightness in the two \textit{Planck}/HFI bands of
interest.

The SPIRE observations were carried out by mapping the field with a
series of scans in one (the nominal) orientation and then a second
time (with a different obsid) with a nearly orthogonal
orientation (the cross-scan).  Maps were made by processing simultaneously the data
from scans in both the nominal and orthogonal orientations.
For each field, we used the Level 0 SPIRE timelines
products from the HSA. The observations are then processed with {\tt
HIPE} to produce PMW and PLW maps with pixel sizes 10\arcs\ and
14\arcs, respectively.
Destriping was applied using the \Herschel/{\tt HIPE} destriper
module.  A constant offset was also removed that is equal to the median flux
level of the map, because SPIRE (like HFI) only measures relative
fluxes, but this is of no consequence for the relative gain
comparison.

The HFI maps of the same fields were produced as cutouts of the 545
and 857\GHz\ whole-sky {\tt HEALPix}\footnote{
See \citet{gorski2005} and \url{http://healpix.sf.net}}
maps, corrected for zodiacal light, from the 2015 \textit{Planck} data
release \citep[PR2;][]{planck2014-a09}. We reprojected them on the
SPIRE maps coordinates with a pixel size of $3\arcm\ \times 3\arcm$.

Because the SPIRE beams are much smaller than those of HFI, we
convolved the SPIRE maps with the HFI effective Gaussian beams
(Sect.\,\ref{sect:calib}).  We checked that the approximation
of ignoring slight beam variations over the sky has a negligible
impact on our analysis. 
Finally, we interpolated the convolved SPIRE maps to the HFI pixel grid.

\subsection{Estimated noise maps} 
\label{ssect:quala}

To estimate the high-frequency statistical noise for the HFI
observations
\bfc{we took advantage of the redundancy of \textit{Planck} observations during a stable pointing period (referred to as a ring; see \citealp{planck2013-p03f}).  Among the products in the PR2 release there are two independent HFI maps produced by splitting each ring into two
equal duration parts, the so-called half-ring maps (\citealp{planck2014-a09}, Appendix A.1).
The noise is assessed} using the difference of cutouts from the two half-ring
maps, following the method described in
\citet{2002A&A...393..749M} and \citet{IRIS2005} in which the noise is
assumed to be stationary.  
The histogram of the brightnesses in these
difference maps is very well described by a Gaussian in all the fields
(see, e.g., the fit in Fig.\,\ref{fig:diff_maps}, lower left, for the
HFI Polaris field).  The final estimated noise map (e.g.,
Fig.\,\ref{fig:diff_maps}, upper left) is weighted using the coverage.
We note that this estimate is consistent with realizations based on
the PR2 variance map \citep{planck2014-a09}.

\begin{figure}
\begin{center}
\begin{tabular}{cc}
\includegraphics[width=8cm]{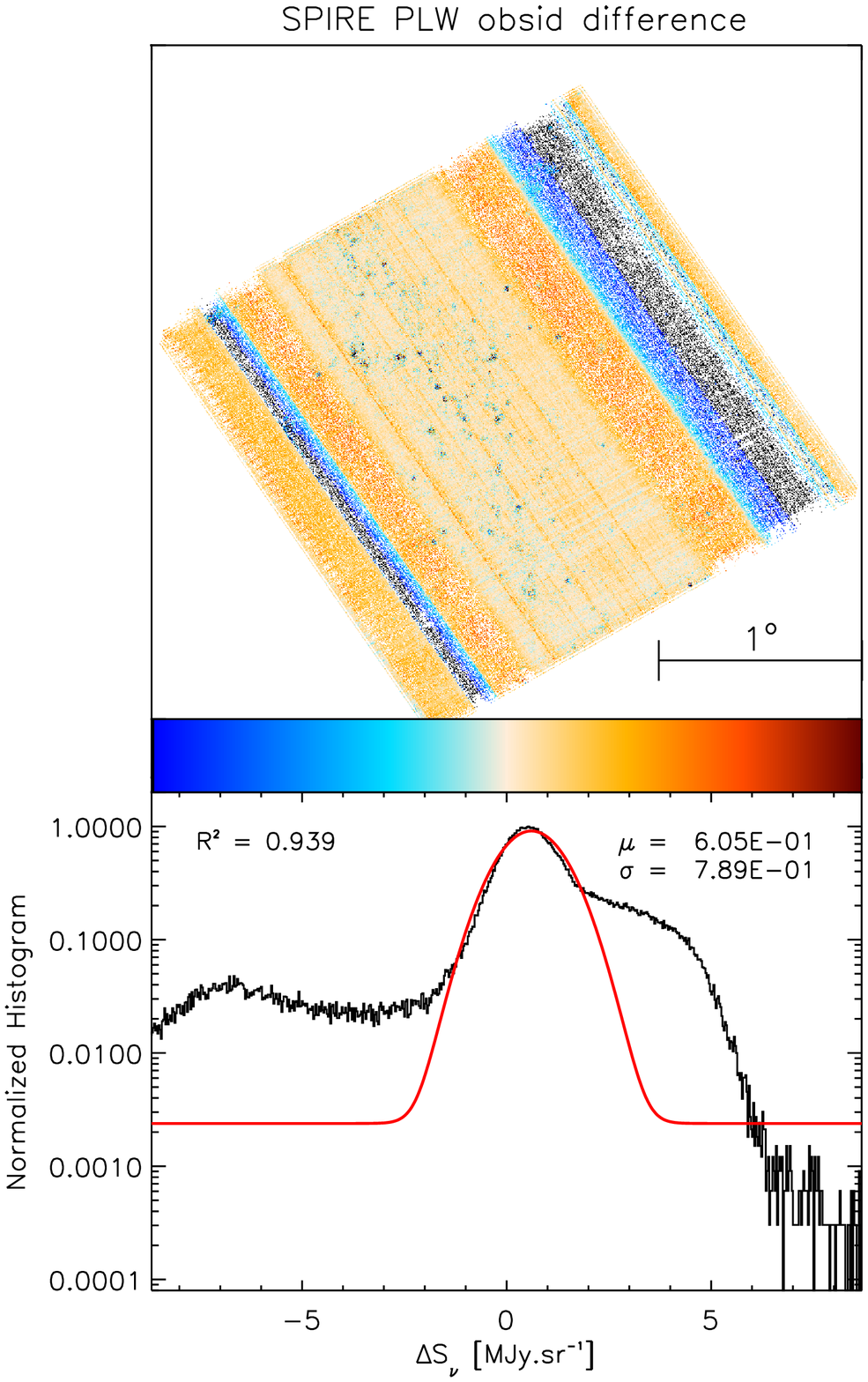}
\end{tabular}
\vspace{-.8cm}
\caption{Same as Fig.~\ref{fig:diff_maps} but for a SPIRE difference
map dominated by systematic effects (Hi-GAL field 37\_0, 
\bfc{$2\pdeg5$ square}). Although these fields might
potentially be recovered using specifically tuned pipelines, we
decided to discard them for reasons of consistency and accuracy.}
\label{fig:bad_diff_maps}
\end{center}
\end{figure}

The procedure for SPIRE was similar.  We made separate maps from data
for scans in each of the two orthogonal orientations (the two separate
obsids) separately, creating a difference map that was then
weighted by the coverage.  In some SPIRE fields the histograms for the
difference maps showed distributions that are slightly broader than the central
Gaussian core with low amplitude tails at high and low values of
brightness difference.  These are mostly due to $1/f$ noise (see also
Sect.\,\ref{ssect:imple} below) that is not perfectly removed when a
map is produced from data in only a single scan orientation, as
opposed to both (this is the fundamental reason for this observational
strategy).  Varying spatial coverage between the orthogonal sets of
observations, which is especially apparent along the map edges, also
contributes to broadening the histogram.  However these effects,
which are already small at the SPIRE angular resolution, become completely
negligible when the fields are brought to the HFI angular resolution.
Similarly, the values shown here for the SPIRE difference maps are not
directly comparable to the noise estimates discussed in
Sect.\,\ref{ssect:corr} because the HFI 545 and 857\GHz\ beams
encompass roughly 450 and 870 pixels in the corresponding SPIRE map,
respectively.  At the same resolution and pixelization the
distribution of SPIRE noise has a much smaller variance than is found
for the noise in the HFI map.

The SPIRE Polaris field (Fig.\,\ref{fig:diff_maps}, right, and
Fig.~\ref{fig:pwsp_polaris} below) is a special case, since it is the result
of mapping two overlapping subfields\footnote{
The structural and statistical properties of the subfield extending to
the upper left were studied by \citet{Miville-Deschenes:2010uq} using
a power spectrum analysis.}
(hence two pairs obsids), and the effect of weighting by the
coverage is clearly seen in the estimated noise map.

Finally, we note that several additional fields were initially part of
the analysis but were discarded because they displayed strong residual
effects. An example of such a field is shown in
Fig.\,\ref{fig:bad_diff_maps}.  As a result of these residual systematic
effects, adding such fields would not improve the accuracy of the
joint estimate of the cross-calibration relative gain. Similarly,
lower brightness fields are not useful because the instrumental noise
limits the accuracy of the comparison.

\section{Bandpass corrections}
\label{sect:ccolor}

To compare flux density or brightness measurements from two different
instruments, it is necessary to apply a correction that takes the differences in 
their respective net spectral responses into account. This bandpass
correction should not to be confused with the standard colour
correction that, for a given bandpass, converts the brightness
obtained at the nominal frequency from a given SED to that from
another SED. A bandpass correction converts, for a given SED, the
brightness obtained at the nominal frequency from a given bandpass to
the brightness that would be obtained from another bandpass.
\bfc{Because the SED varies both within a given field and from field to field, bandpass corrections are essential in preparing the data for assessment of the cross-calibration relative gain; a corollary is that there is no unique bandpass conversion factor between the uncorrected HFI and SPIRE data.}

Although the two pairs of HFI and SPIRE bandpasses are similar, they are nevertheless
sufficiently different that bandpass corrections are required,
especially in the case of the 545\GHz\ and 500\microm\ bandpasses which
overlap by only about two-thirds (see Fig.\,\ref{fig:rsrfs}).
\bfc{Only after the bandpass correction has been applied can one compare the two independent brightness measurements to investigate whether the absolute calibrations are consistent within their uncertainties.  Technically, this comparison gives the desired measurement of the cross-calibration relative gain, which is expected to be independent of the SED and field on the sky.}

\begin{figure}
\begin{center}
\includegraphics[width=9 cm]{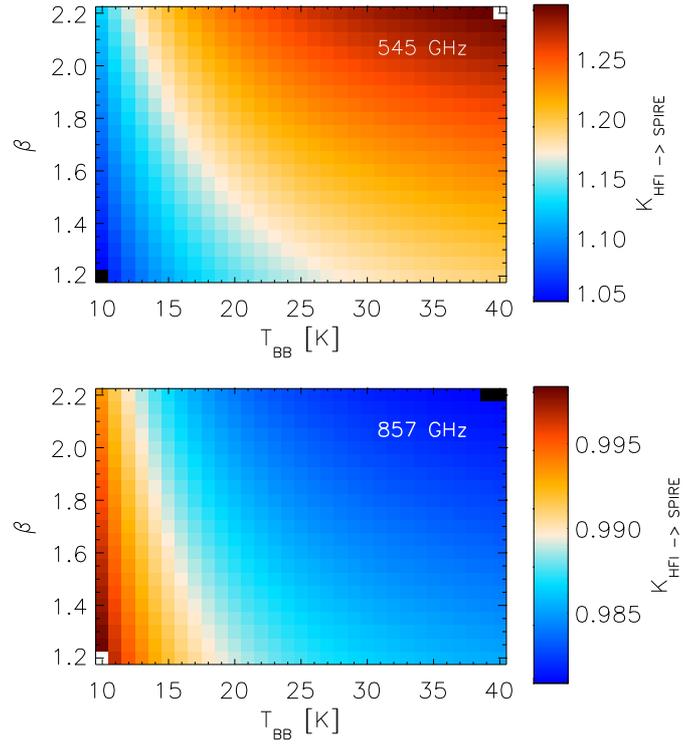}
\caption{Correction factors from measurements with HFI to what would
be observed by SPIRE, $K_{\text{HFI}->\text{SPIRE}}$
(Eq.\,\ref{eq:ccfactor}), for MBB spectra (Eq.\,\ref{eq:greybody})
with a range of temperatures, $T_{BB}$, and emissivity indices,
$\beta$, at the two HFI frequencies. We decreased the
resolution along the temperature axis by a factor of
10 compared to the tabulated values to a step of 1~K.  The
anti-correlation between $T_{BB}$ and $\beta$ 
can produce the same MBB SED shape across the bandpass and, hence, the same
correction.
}
\label{fig:ccmaps}
\end{center}
\end{figure}

\begin{figure}
\begin{center}
\includegraphics[width=8 cm]{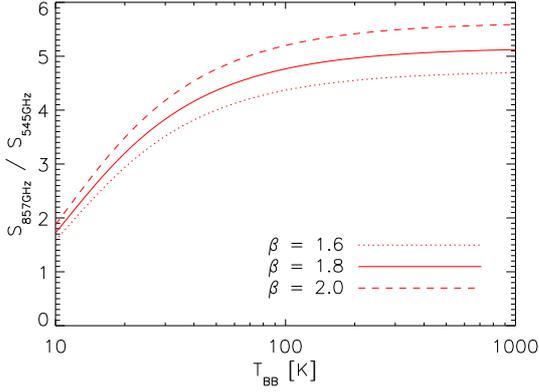}
\caption{\textit{Planck} HFI 857 to 545\GHz\ brightness ratio versus MBB
spectrum temperature, $T_{BB}$, for three different values of the
emissivity index ($\beta = 1.6$, $1.8$, and $2.0$). For our
application, dust temperatures would be in the low end of the range,
allowing some discrimination of $T_{BB}$.}
\label{fig:slope2temp}
\end{center}
\end{figure}

\begin{figure}
\begin{center}
\includegraphics[width=9 cm]{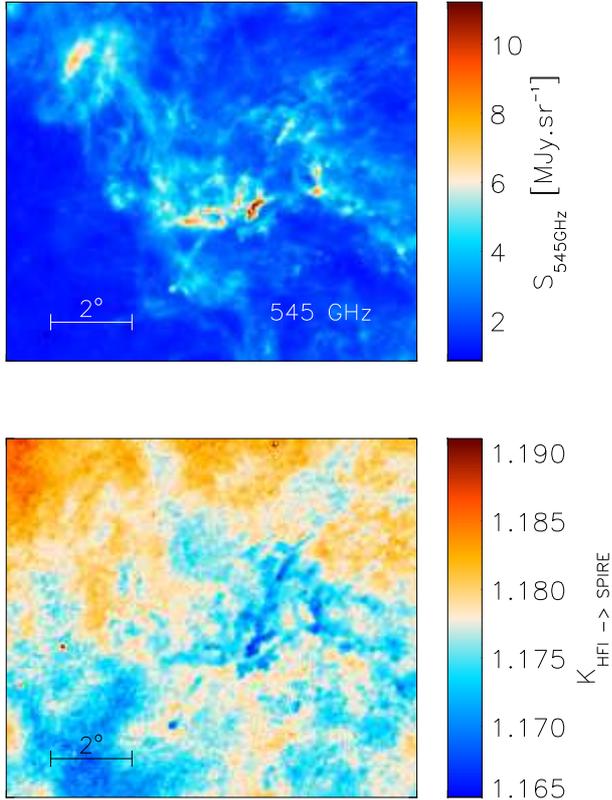}
\vspace{-1.5cm}
\caption{{\it Top:} HFI 545\GHz\ 
\bfc{spanning} the Polaris field. 
{\it Bottom:} $K_{\text{HFI}->\text{SPIRE}}$ bandpass correction,
$K_{545}$ at 545\GHz\ (Eq.~\ref{eq_cc}), for the same field.
}
\label{fig:polaris_ccmap}
\end{center}
\end{figure}

\subsection{Definition} 
\label{sect:cc}

As described in Sect.\,\ref{sect:calib}, \Herschel/SPIRE and
\textit{Planck}/HFI each provides maps of monochromatic surface brightnesses
$\tilde{S}_{\nu_0}$, at their individual reference frequency 
$\nu_0$, assuming a reference spectrum $\nu\ S_{\text{ref}}(\nu) =
\text{constant}$ (see Eq.\,\ref{eq:snu0}).
We need a bandpass correction $K_{\text{HFI}->\text{SPIRE}}$ to
convert the observed brightness from HFI (at
$\nu_{0_\text{H}}$) to that which would be observed in the
paired SPIRE bandpass (at $\nu_{0_\text{S}}$), so that the
bandpass-corrected HFI map can be compared to the SPIRE map.  Thus
\begin{equation}
K_{\text{HFI}->\text{SPIRE}} = 
\cfrac{\nu_{0_\text{H}}}{\nu_{0_\text{S}}} 
\cfrac{\int \cfrac{\rsf_\text{H}(\nu)}{\nu}\,d\nu}{\int \cfrac{\rsf_\text{S}(\nu)}{\nu}\,d\nu} 
\cfrac{\int \rsf_\text{S}(\nu)S_\nu\,d\nu}{\int \rsf_\text{H}(\nu)S_\nu\,d\nu} \, ,
\label{eq:ccfactor}
\end{equation}
where $\rsf_\text{H}$ and $\rsf_\text{S}$ are the net spectral responses for extended
emission for HFI and SPIRE, respectively (as shown in Fig.\,\ref{fig:rsrfs}).

\subsection{Evaluating the bandpass correction}
\label{ssect:essp}

Equation\,\eqref{eq:ccfactor} shows that
$K_{\text{HFI}->\text{SPIRE}}$ depends on the shape of the source SED,
$S_\nu$. In rare cases $S_\nu$ is well known beforehand (e.g., for a
calibration source) but in general its shape has to be estimated (or
assumed).

A simple and robust model of the SED of thermal emission by Galactic
dust in the submillimetre range is a modified blackbody (MBB) spectrum
with a temperature, $T_{BB}$, and emissivity index, $\beta$, in the
form
\begin{equation}
S_\nu \propto B_\nu(T_{BB}) \times \nu^{\beta} \, .
\label{eq:greybody}
\end{equation}
Typical $T_{BB}$ and $\beta$ values for diffuse dust emission, which
is the dominant component in the selected maps, range from 10 to 40~K
and 1.4 to 2.5, respectively \citep[e.g.][]{planck2013-p06b, Paradis:2010kx}.

We used Eqs.~\eqref{eq:ccfactor} and \eqref{eq:greybody} to produce
tables of the $K_{\text{HFI}->\text{SPIRE}}$ bandpass-correction
factors
\begin{equation}
\label{eq_cc}
K_{545} = \frac{\tilde{S}_{PLW}}{\tilde{S}_{545\,\text{GHz}}} \ \ \text{and} \ \ K_{857} = \frac{\tilde{S}_{PMW}}{\tilde{S}_{857\,\text{GHz}}}  \, 
\end{equation}
for a range of $T_{BB}$ from 10 to 40~K with a step of 0.1~K, and a
range of $\beta$ from 1.2 to 2.2 with a step of $0.05$.  These are
shown in Fig.\,\ref{fig:ccmaps}.  

Several approaches can be used to estimate $T_{BB}$ and $\beta$ from
the data, one of which is to fix the emissivity index at an
appropriate value deduced from previous studies and deduce the
temperature from the measured ratio \textit{Planck} HFI 857 to 545\GHz\
brightness ratio (Fig.\,\ref{fig:slope2temp}).  This ratio could be
obtained from standard linear regression between the two entire maps,
but this would not take local temperature variations into account, and these variations 
can be important.
Alternatively, the temperature can be deduced from the brightness ratio
at each pixel.  This is implemented in the {\tt HIPE} pipeline
developed to determine SPIRE maps offsets, with 8\arcmin-resolution
\textit{Planck} maps.

We implemented a second option, exploiting full-sky maps of $T_{BB}$
and $\beta$ produced from multi-frequency data as part of the \textit{Planck}
products \citep{planck2013-p06b, planck2013-p28thermaldust}.  
Neither $T_{BB}$ and $\beta$ are individually important as long as,
together in Eq.\,\eqref{eq:greybody}, they describe the slope of the
SED that are fit in the data.  Thus at every HFI pixel we can readily
estimate the bandpass correction by interpolation in the above tables
of $K$, producing bandpass-correction maps.  
An example is shown in
Fig.\,\ref{fig:polaris_ccmap} for the Polaris field at 545\GHz.
We note that the bandpass corrections are different from pixel to pixel, 
with a range that is large compared to the uncertainty that we find for the cross-calibration relative gain below.  
Thus accurately computing and applying bandpass corrections is essential.

\section{Pixel-to-pixel comparison}
\label{sect:pixtopix}

\begin{figure*}
\begin{center}
\includegraphics[width=16 cm]{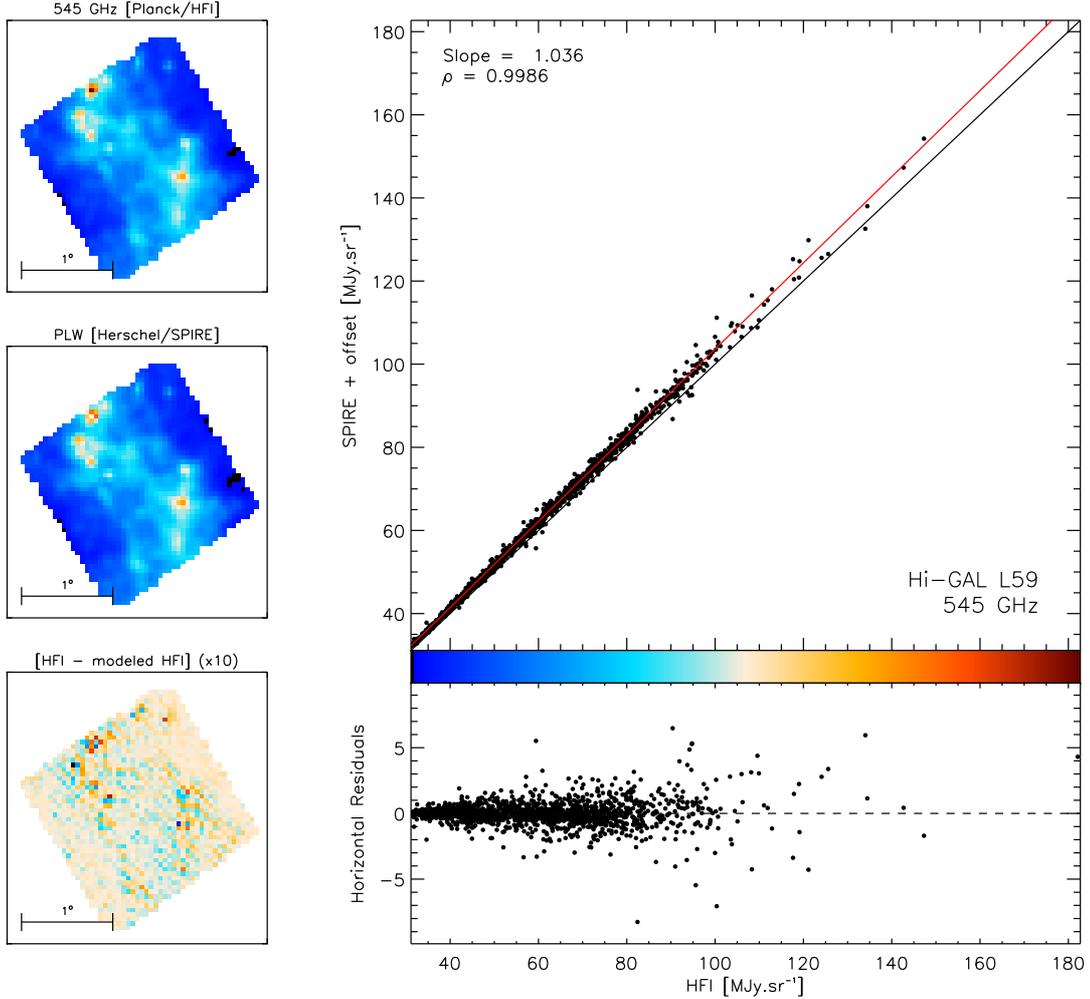}
\vspace{-0.8cm}
\caption{Correlation and pixel-to-pixel comparison between HFI and
SPIRE observations of the Hi-GAL \lfn\ field at 545\GHz.
{\it Top left:} HFI and SPIRE maps of the \lfn\ field. 
{\it Top right:} SPIRE--HFI pixel-to-pixel correlation. The linear fit
from the unweighted ``X on Y'' regression is plotted in red,
with the 1-to-1 line in black for reference.  The offset removed from
the SPIRE map during the map-making has been obtained from the fit and
added back.
{\it Bottom right:} Horizontal residuals from the fit, as a function of brightness.
{\it Bottom left:} Map of the horizontal residuals, with
a colour range that is ten times finer (centred on 0), to enhance the detail.
}
\label{fig:xc_higal}
\end{center}
\end{figure*}

\bfc{After bandpass correction with
$K_{\text{HFI}->\text{SPIRE}}$ an observation of a field with HFI} should yield brightnesses equal to
those measured with SPIRE (but the latter is missing an offset). This
appears to be the case, as illustrated in Fig.\,\ref{fig:xc_higal}.
Any difference in slope compared to unity should be consistent with
(or lower than) the combination of reported absolute calibration
errors.

\subsection{Degree of correlation}
\label{ssect:corr}

The Pearson correlation coefficient between the pixel brightnesses in
the two maps ($S_\nu^{\text{i}}$ in map i, in \MJysr) is the ratio of
their covariance to the product of their standard deviations
$\sigma_i$ (not to be confused with statistical uncertainties), i.e.
\begin{equation}
\rho_\text{\sc hfi--spire} = \frac{\text{Cov}\,(S_\nu^{\text{\sc hfi}} \,,\, S_\nu^{\text{\sc spire}})}{\sigma_{\text{\sc hfi}} \ \sigma_{\text{\sc spire}}} \,,
\end{equation}
which  is unaffected by potential difference between the two
calibrations.  Therefore, we use the Pearson correlation coefficient to
assess the degree of linear correlation between the SPIRE and HFI
surface brightnesses in a map.  In the presence of noisy variables,
the computed correlation coefficient, $\hat{\rho}_\text{\sc
  hfi--spire}$, always underestimates the true value (see
Sect.\,\ref{ssect:compare}).

Estimated correlation coefficients for each field at the two frequency
bands are reported in Table\,\ref{tab:cmpresults}. They average to
0.9986 ($\pm\, 0.08\,\%$) and 0.9985 ($\pm\, 0.07\,\%$)
at 545 and 857\GHz, respectively. This high degree of linear
correlation between the SPIRE and HFI signals, although expected,
confirms that the SPIRE\,/\,HFI relative gain does not depend on
brightness and is a useful validation that a simple linear fit
procedure can be used to derive the cross-calibration relative gain
and the SPIRE map offsets (Sect.\,\ref{ssect:gains}).

\subsection{Relative gain measurements}
\label{ssect:gains}

We used scatter plots to estimate the relative gain
($G^{\text{Scat}}$) between the two instruments.  Following the
hypothesis of linearity as demonstrated in Sect.\,\ref{ssect:corr},
the relative gain is the slope of the correlation, independent of the
zero levels.
However, the same procedure can be used to find and restore the offset
that has been removed from the SPIRE map during the map-making.
Thus, the relationship is
\begin{equation}
S_\nu^\text{\sc spire} = G^{\text{Scat}} \times S_\nu^\text{\sc hfi} - \text{O} \, .
\label{eqn:gainfit}
\end{equation}
Because the noise in the HFI map is considerably larger than in the
convolved SPIRE map (Sect.~\ref{ssect:quala}), in practice we carried
out the unweighted fit to the data in the sense of a regression of ``X
on Y'' rather than ``Y on X'' and then inverted the slope to correspond
to $G^{\text{Scat}}$ as written in Eq.\,\eqref{eqn:gainfit}.

An example of the SPIRE--HFI scatter plot and computed linear fit is
shown in Fig.\,\ref{fig:xc_higal} for the \lfn\ field of the Hi-GAL
survey. This analysis has been carried out for the 15 selected
fields for the two frequency bands. Results for the relative gains,
the slopes of the linear fits, are given in
Table\,\ref{tab:cmpresults}.  The average and standard deviation of
the estimated values of $G^{\text{Scat}}$ is \gcinq\,\egcinq\ and
\ghuit\,\eghuit\ at 545 and 857\GHz, respectively.

\begin{table*}
\begingroup 
\newdimen\tblskip \tblskip=5pt
\caption{Estimates of the relative gains, $G^\text{Scat}$, and Pearson
correlation coefficients, $\hat{\rho}_\text{HFI--SPIRE}$, at 545 and
857\GHz.} 
\label{tab:cmpresults}
\vskip -6mm
\footnotesize 
\setbox\tablebox=\vbox{
\newdimen\digitwidth
\setbox0=\hbox{\rm 0}
\digitwidth=\wd0
\catcode`*=\active
\def*{\kern\digitwidth}
\newdimen\signwidth
\setbox0=\hbox{+}
\signwidth=\wd0
\catcode`!=\active
\def!{\kern\signwidth}
\newdimen\decimalwidth
\setbox0=\hbox{.}
\decimalwidth=\wd0
\catcode`@=\active
\def@{\kern\signwidth}
\halign{ \hbox to 1.3in{#\leaderfil}\tabskip=1em& 
    \hfil#\hfil\tabskip=2em& 
    \hfil#\hfil& 
    \hfil#\hfil& 
    \hfil#\hfil& 
    \hfil#\hfil& 
    \hfil#\hfil \tabskip=0pt\cr
\noalign{\doubleline}
\omit Name & $G^\text{Scat}_\text{545}$ & $G^\text{Scat}_\text{857}$ & $\hat{\rho}_\text{545}$ &$\hat{\rho}_\text{857}$ & $\Delta u^\text{545}$ & $\Delta u^\text{857}$ \cr
\omit &  &  & & & \% of $u_\rho^{545}$ & \% of  $u_\rho^{857}$ \cr
\noalign{\vskip 3pt\hrule\vskip 5pt}
Hi-GAL \lfn & 1.036 & 0.997 & 0.9986 & 0.9982 & 0.14 & 0.18 \cr
Hi-GAL \lth & 1.054 & 1.002 & 0.9989 & 0.9993 & 0.11 & 0.07 \cr
Hi-GAL Field 0\_0 & 1.058 & 1.021 & 0.9984 & 0.9976 & 0.16 & 0.24 \cr
Hi-GAL Field 2\_0 & 1.046 & 1.007 & 0.9992 & 0.9992 & 0.08 & 0.08 \cr
Hi-GAL Field 4\_0 & 1.039 & 0.987 & 0.9985 & 0.9983 & 0.15 & 0.17 \cr
Hi-GAL Field 6\_0 & 1.048 & 1.006 & 0.9988 & 0.9990 & 0.12 & 0.10 \cr
Hi-GAL Field 8\_0 & 1.047 & 1.005 & 0.9985 & 0.9984 & 0.15 & 0.16 \cr
Hi-GAL Field 33\_0 & 1.047 & 1.000 & 0.9993 & 0.9991 & 0.07 & 0.09 \cr
Hi-GAL Field 35\_0 & 1.045 & 1.004 & 0.9987 & 0.9983 & 0.14 & 0.17 \cr
Hi-GAL Field 39\_0 & 1.053 & 1.005 & 0.9995 & 0.9994 & 0.06 & 0.06 \cr
Hi-GAL Field 332\_0 & 1.052 & 1.006 & 0.9981 & 0.9970 & 0.19 & 0.30 \cr
Hi-GAL Field 334\_0 & 1.048 & 1.005 & 0.9989 & 0.9991 & 0.11 & 0.09 \cr
Aquila & 1.047 & 0.997 & 0.9988 & 0.9982 & 0.13 & 0.19 \cr
Polaris & 1.030 & 0.989 & 0.9983 & 0.9981 & 0.17 & 0.19 \cr
Spider & 1.049 & 1.010 & 0.9958 & 0.9978 & 0.43 & 0.22 \cr
\noalign{\vskip 5pt\hrule\vskip 3pt}
Average & \gcinq\,\egcinq & \ghuit\,\eghuit & 0.9986 ($\pm\, 0.08\,\%$) & 0.9985 ($\pm\, 0.07\,\%$) & \dots & \dots  \cr
\noalign{\vskip 5pt\hrule\vskip 3pt}
}}
\endPlancktable 
Notes:\par
Averages (and standard deviations) are over all fields, calculated a
posteriori.\par
Last two columns show the difference, in percentage, between the two
HFI uncertainty estimates, as described in Sect.\,\ref{sect:pixtopix},
which highlights that our assumptions about the HFI uncertainties and
the linearity between instrument measurements are very reasonable.\par
\endgroup
\end{table*} 

\subsection{Comparison of estimates of the uncertainty in the HFI maps}
\label{ssect:compare}

Considering that the uncertainties in the convolved SPIRE maps are
very small in comparison to the uncertainties in the HFI maps, we can
attribute the dispersion about the regression line to the HFI
uncertainties alone.  We can estimate the $1\sigma$ uncertainty,
$u_\text{HFI}^\text{fit}$, from the dispersion of the residuals in the
horizontal direction:
$S_\nu^\text{\sc hfi} - (S_\nu^\text{\sc spire} +   \text{O} )/G^{\text{Scat}}$.
For the example \lfn\ field, these residuals are shown in the lower
panels of Fig.\,\ref{fig:xc_higal}.

Under the assumption of Gaussianity of the noise in our maps, which is
reasonable here as shown in Sect.\,\ref{ssect:quala}, the Pearson
correlation coefficient allows us to make an alternative estimate, a
posteriori, of the noise in the HFI maps.  Assuming that the measured
signal can be written as $S_\text{meas} = S_\text{true} +
S_\text{noise}$, which is a reasonable assumption, the computed
correlation coefficient reads
\begin{equation}
\hat{\rho}_\text{\sc hfi--spire}  = \left[ \left(1 + \frac{u_{\text{\sc hfi}}^2}{\sigma_{\text{\sc hfi}}^2} \right) 
\ \left(1 + \frac{u_{\text{\sc spire}}^2}{\sigma_{\text{\sc spire}}^2} \right) \right]^{-1/2} 
\ \rho_\text{\sc hfi--spire}  \, ,
\label{eqn:rho_tilde}
\end{equation}
where $u_\text{i}$ are $1\sigma$ uncertainties of the brightnesses
characterizing $S_\text{noise}$.
In our selected fields, the signal variance $\sigma_\text{HFI}^2 \simeq
\sigma_\text{SPIRE}^2$ is much higher than that of the uncertainty,
i.e. $u_i^2 \ll \sigma_i^2$ in both SPIRE and HFI maps. Furthermore,
as discussed above, $u_\text{SPIRE}^2 \ll u_\text{HFI}^2$.  Thus
Eq.~\eqref{eqn:rho_tilde} simplifies to
\begin{equation}
\hat{\rho}_\text{\sc hfi--spire}  = \left(1 + \frac{u_{\text{\sc hfi}}^2}{\sigma_{\text{\sc hfi}}^2} \right)^{-1/2} 
\ \rho_\text{\sc hfi--spire}  \, .
\label{eqn:rho_tilde_simple}
\end{equation}
Because $u_\text{HFI}^2 \ll \sigma_\text{HFI}^2$, we can use the
variance of the measured HFI brightness as a first order estimate of
$\sigma_\text{HFI}^2$ (the variance of the true, noiseless signal).
Furthermore, we can assume that $\rho_\text{\sc hfi--spire} = 1$ on the RHS.
Using the computed Pearson correlation coefficients $\hat{\rho}$ in
Table\,\ref{tab:cmpresults} on the LHS, we solve
Eq.\,\eqref{eqn:rho_tilde_simple} for $u_\text{HFI}$, designating this
a posteriori estimate as $u_\text{HFI}^{\, \rho}$.

In the last two columns (for the two frequency bands) of
Table\,\ref{tab:cmpresults}, we tabulate the percentage
difference of these two estimates for each field:
$100 \, | (u_\text{HFI}^\text{fit} - u_\text{HFI}^{\,\rho})/u_\text{HFI}^{\, \rho} |$.
As expected, we find these to be in very good agreement (better than
0.4\,\%). The overall consistency further demonstrates that the gain
does not depend on brightness and is consistent with the HFI
brightness uncertainties dominating over those of SPIRE.

\subsection{Contributions to the error budget for $G^{\text{Scat}}$}
\label{ssect:ebud}

The relative gains in Table~\ref{tab:cmpresults} show that at 857\GHz\
the SPIRE and HFI photometric calibrations agree to within 1\,\%.
However, at 545\GHz\ there is a 4.7\,\% discrepancy that is
statistically significant ($6.8\sigma$).
Each instrument has absolute calibration errors that were discussed in Sect.\,\ref{sect:calib}.
As we showed in Appendix\,\ref{sect:neptmod}, the Neptune models used for
calibration of the two instruments are very close and can only explain at most a 0.31\,\%
departure from unity. Furthermore, the statistical measurement error
from each instrument combined quadratically amount to an uncertainty
of 1.86\,\% and 2.06\,\% in the relative gain at 545 and
857\GHz. However, the SPIRE effective beam solid angle has an uncertainty of
4\,\% \citep{Griffin:2013vn}, acting as a systematic effect on the
SPIRE photometric calibration and thus on the gain estimates as
well. This dominates the \textit{Planck}/\Herschel\ cross-calibration error
budget presented here and could account for the observed 4.7\,\%
discrepancy at 545\GHz; the potential revision of the value of the solid angle discussed in Appendix\,\ref{sect:spicalib} would reduce the discrepancy to 2.7\,\%.

\begin{figure*}
\begin{center}
\begin{tabular}{c}
\includegraphics[width=16cm]{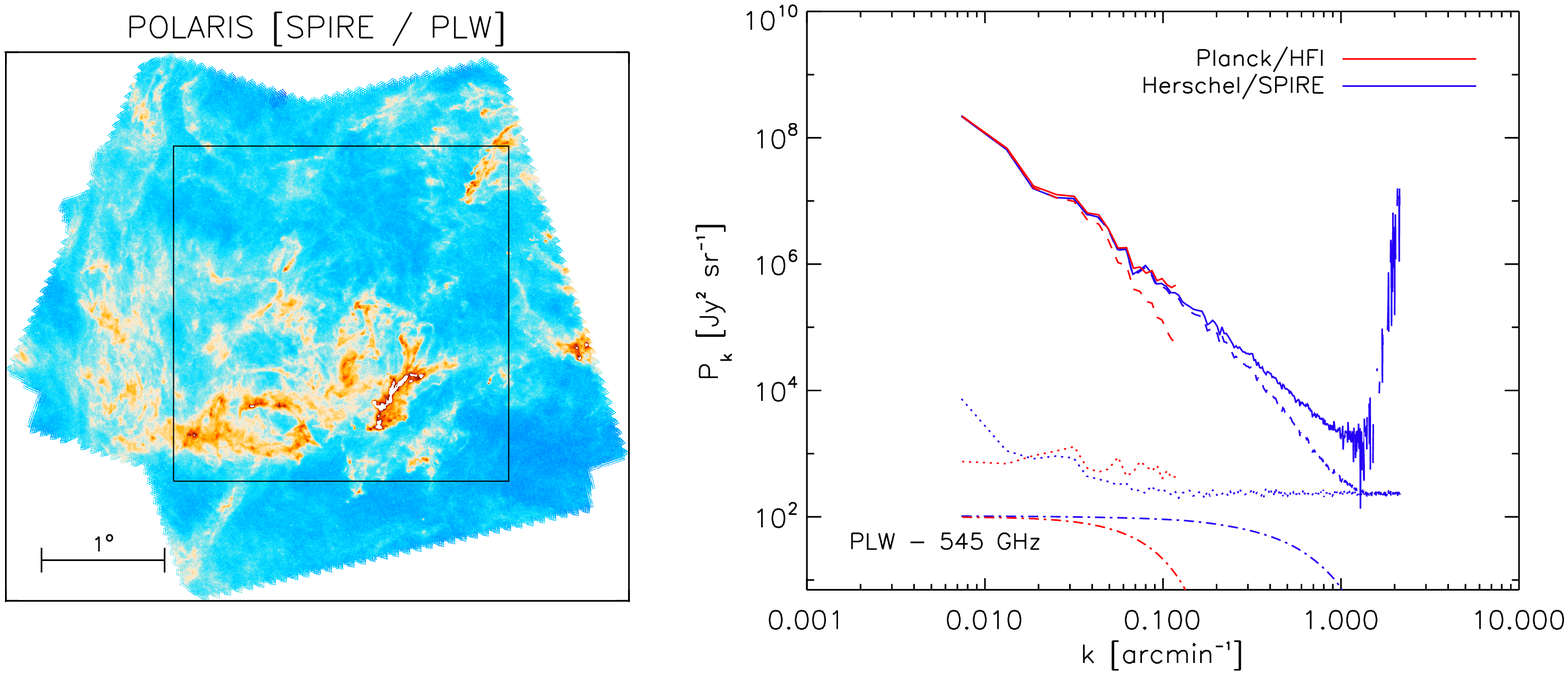}\\
\includegraphics[width=16cm]{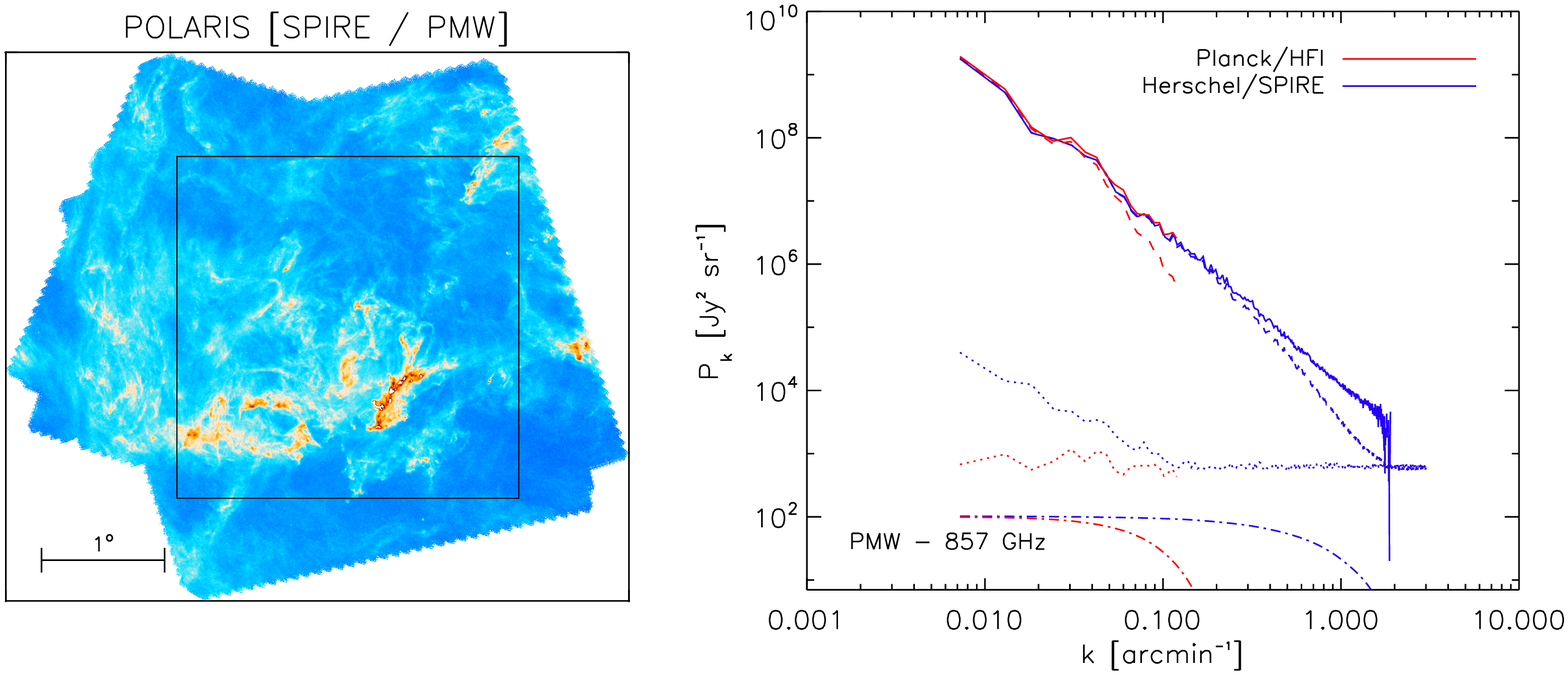} \\
\end{tabular}
\caption{
{\it Left:} SPIRE PLW and PMW maps of the Polaris field. The black
square outlines the region of the map that is used to compute the
power spectra shown on the right.
{\it Right:} Power spectra of SPIRE map (blue) and bandpass-corrected
HFI (red) map for PLW (top) and PMW (bottom) in the Polaris field. The
solid lines are the dust power spectra obtained from the image power
spectra (dashed line), from which the noise spectra (dotted line) are subtracted
followed by division by the effective beam power spectra (dash dotted line; arbitrarily
shifted on the y-axis).
}
\label{fig:pwsp_polaris}
\end{center}
\end{figure*}

\section{Cross calibration with power spectrum measurements}
\label{sect:powspec}

Power spectra offer a means of probing the spatial structure of
diffuse emission. In our context, we can use power spectra to search
for any scale dependence of the relative gain. This scale dependence
is not expected, but can be introduced artificially by the data
processing (any filtering process and the map-making). It is therefore
of particular interest to check the level of agreement between power
spectra of bandpass-corrected maps produced by \textit{Planck}/HFI and those
from \Herschel/SPIRE.

However, the HFI and SPIRE spatial resolutions differ greatly and that
of HFI limits the largest spatial frequency (lowest spatial scale) for
comparison to about $k=0.3\,\text{arcmin}^{-1}$.  As a result of map
pixelization effects, we actually limit our power spectrum comparison
to $k < 0.1\,\text{arcmin}^{-1}$.  Furthermore, the SPIRE fields are
limited in sky coverage, which limits the smallest spatial frequency
to about $k=0.007\,\text{arcmin}^{-1}$(proportionately lower for the
three larger fields).  At those scales, variations in the power
spectra are dominated by cosmic variance but because cosmic variance
is the same for the two maps it need not be considered in the
comparison.  This is not a large range in scale but thanks to the
number of fields the power spectrum comparison still allows us to
check for any systematic scale dependence of the gain.

\subsection{Implementation}
\label{ssect:imple}

We implement the estimation of the power spectrum based on the
methodology presented in \cite{2002A&A...393..749M} via the IDL {\tt
  FFT} routine.  In overview, the power spectrum of an image $S_{x,y}$
whose Fourier transform is $\widetilde{S}_{k_x,k_y}$ is computed from the
map of the amplitude $A_{k_x,k_y}$ defined by
\begin{equation}
A_{k_x,k_y} = \widetilde{S}_{k_x,k_y}\ \widetilde{S}_{k_x,k_y}^\star = |\, \widetilde{S}_{k_x,k_y}  |^2 \, .
\label{eqn:ak}
\end{equation}
The power spectrum $P(k)\,dk$ is an angular average of $A_{k_x,k_y}$
between $k$ and $k+dk,$ where $k=\sqrt{k_x^2+k_y^2}$.

To compare the power spectra of the dust signal we need to take the noise contamination and the effect of beam
convolution into account. Formally,
\begin{equation}
P_\text{signal}(k) = \frac{P(k) - b_n(k)}{B(k)} \, ,
\end{equation}
where $b_n(k)$ is the noise power spectrum and $B(k)$ the beam power
spectrum. The noise power spectrum is computed from the estimated
noise maps produced from redundant observations, as described in
Sect.\,\ref{ssect:quala}.
As mentioned, there is a $1/f$~noise component in the SPIRE estimate,
which is clearly visible in the noise power spectra in
Fig.\,\ref{fig:pwsp_polaris}. This component is not in the SPIRE power
spectra computed on the total map, but because this component is about
20 -- 50 times smaller than signal power spectrum at low $k$ and it
does not impact our analysis.

In Fig.\,\ref{fig:pwsp_polaris} we show how the deconvolved SPIRE
power spectrum at full resolution compares to that from HFI. This
comparison is qualitatively interesting; as in the pixel-to-pixel
comparison, we see a very good agreement across the range of spatial
frequencies common to both instruments.
Furthermore, the same power-law exponent appears to apply at the
higher spatial frequencies accessible only to SPIRE, which is consistent with
the study by \citet{Miville-Deschenes:2010uq}.

\begin{figure*}
\begin{center}
\includegraphics[width=16 cm]{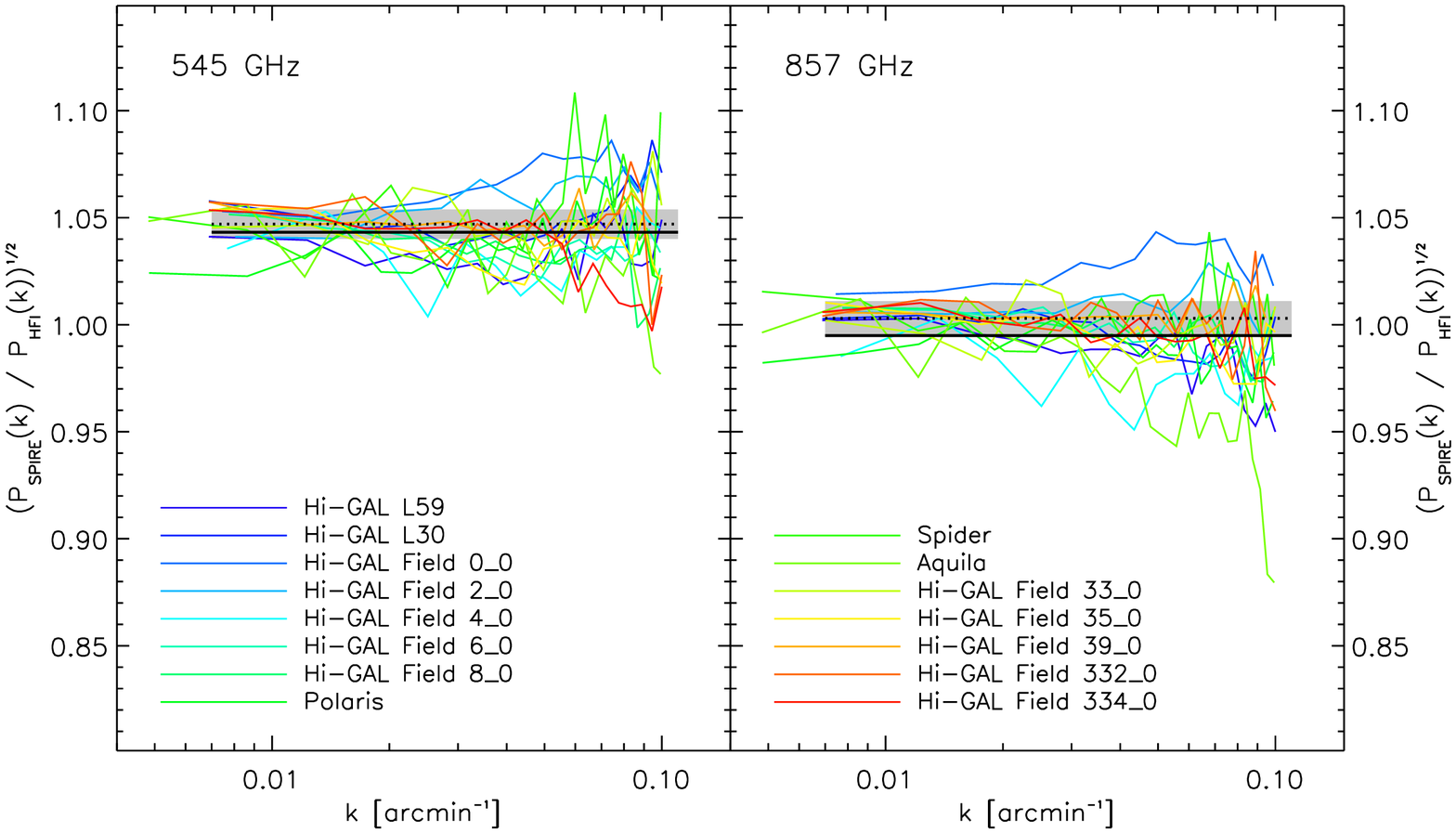}
\caption{Relative gain, from the square root of the SPIRE-to-HFI power
spectrum ratio, for each field at 545\GHz\ (left) and 857\GHz\
(right). The mean gain $G^{\text{PSD}}$ from the power spectrum
analysis (solid line) is in close agreement with the mean gain
$G^{\text{Scat}}$ from pixel-to-pixel analysis (dotted line), which is well
within the 1$\sigma$ dispersion of the latter (shaded in grey).
}
\label{fig:rpwsp_all_fields}
\end{center}
\end{figure*}

\subsection{Relative gain from power spectra}
\label{ssect:relgain}

To compare SPIRE and HFI photometric calibrations quantitatively, we
compute $P(k)$ on the bandpass-corrected HFI maps and on the SPIRE
maps brought to the HFI resolution.  This mitigates against errors due
to beam accuracy and differences in pixel window functions.

The {\tt HIPE} pipeline does not optimize the orientation of the final
image to be used in the power spectrum analysis and so our maps
feature zero coverage on the corners. Rather than rotating the WCS,
which would lead to significant distortion at small scales in the
power spectrum, or cropping our data to small full-coverage insets
(which was sufficient for the purposes of
Fig.\,\ref{fig:pwsp_polaris}), which would reduce the range of
overlapping scales, we zero-averaged our images and padded them with
zeros. This masking operation in the image, which corresponds to a
convolution of the true power spectrum in Fourier space, is carried
out on both SPIRE and HFI maps, and so does not affect the power
spectrum ratio and thus the relative gain determination.

The relative gain is the square root of ratio of the two power
spectra, i.e.
\begin{equation}
G^{\text{PSD}}(k) = \sqrt{\frac{P_{\text{SPIRE}}(k)}{P_{\text{HFI}}(k)}} \, .
\label{sgain}
\end{equation}
The relative gain is shown in Fig.\,\ref{fig:rpwsp_all_fields} for
each field, where it is seen that there is good agreement from field
to field.  The average (and rms) $G^{\text{PSD}}$ over the spatial
scales analysed is 1.043 ($\pm\, 0.012$) and 0.995 ($\pm\, 0.015$) at 
545 and 857\GHz, respectively.
These estimates are in very good agreement (better than $1\sigma$)
with the pixel-to-pixel average gains $G^{\text{Scat}}$ in
Table\,\ref{tab:cmpresults}.

To search for any variation with spatial scale, we calculated the
average $G^{\text{PSD}}(k)$ in five bins of $k$ across the range
$0.007$ -- $0.1\,\text{arcmin}^{-1}$.  For both frequencies we found
the relative gains $G^{\text{PSD}}(k)$ to be stable with scale with rms variations lower than about
1.5\,\%.

\section{Conclusion}
\label{conclu}

We performed a detailed comparison of the \Herschel/SPIRE and
\textit{Planck}/HFI absolute photometric calibration.  We produced SPIRE maps for 15 fields that contain
bright diffuse emission using publicly released
\Herschel\ data. These maps were corrected for bandpass
differences, reprocessed to match the HFI angular resolution and
pixelization, and then compared to their \textit{Planck} counterparts from
publicly released data. The comparison was carried out using two
methods: a pixel-to-pixel comparison and a power spectrum analysis. We
discarded several fields that were originally in the study to
limit systematic effects arising from significant SPIRE processing
residuals and to maintain a high signal-to-noise ratio in the maps and
a high accuracy in the comparison.

The pixel-to-pixel comparison revealed the expected very high degree
of linearity between the two datasets, and allowed for a robust
estimate of the relative gain between the two instruments in each of
the overlapping bands. The power spectrum analysis provided gain
estimates with similar values and comparable statistical
uncertainties.
The good agreement of the relative calibration measured on distinct
fields shows robustness to a number of potential systematic errors in
either instrument, including nonlinear response, uncorrected time
variability in the gain, and variation of the effective beam.

We found that the estimated relative gains are well inside the
absolute photometric uncertainties quoted for the two instruments of
about 6.4\,\% and 9.5\,\% for HFI and SPIRE, respectively.  
However, because both instruments use the same planetary calibrator,
that contribution to the relative uncertainty can be reduced to the very small difference between the ESA\,3 and ESA\,4 models of Neptune. 
Then the deviations are comparable to the remaining
uncertainty values of 1.4\,\% and 5.5\,\% for HFI and SPIRE.
The difference in the ESA\,3 and ESA\,4 Neptune models used in the HFI
and SPIRE calibrations, respectively, decreases the two relative gain estimates by
only a very small fraction (0.31\,\% at 545\GHz\ and 0.16\,\% at
857\GHz).

At 545\GHz\ the departure of the relative gain from unity is
statistically significant, which raises the question of whether a
systematic calibration correction should be applied prior to any
comparative analysis between HFI maps at 545\GHz\ and SPIRE maps at
500\microm.  However, this departure could be accounted for by the
4\,\% systematic uncertainty of the SPIRE beam solid angle.  Because
this dominates the error budget of the SPIRE--HFI cross calibration,
we provide here the SPIRE\,/\,HFI relative gains with their explicit
dependence on the SPIRE unit conversion factor $K_{\text{PtoE}}$ as follows:
\begin{equation}
G_{545} = \gcinq\, (\pm\, 0.0069) \times \left( \frac{K_{\text{PtoE}}}{24.039} \right)
\label{g545}
\end{equation} 
and
\begin{equation}
G_{857} = \ghuit\, (\pm\, 0.0080) \times \left( \frac{K_{\text{PtoE}}}{51.799} \right)
\label{g857}
\end{equation}
at 545 and 857\GHz\ (or PLW and PMW), respectively, where the
normalizations of $K_{\text{PtoE}}$ are our
adopted values in \PtoEunits.
These formulae would be useful for any future analysis that might make
use of an updated version of the SPIRE $K_{\text{PtoE}}$ (see e.g. 
Appendix~\ref{sect:spicalib}).

We used two methods of comparison, both based on extended emission,
but did not make a comparison based on point sources.  There are SPIRE
counterparts for many extragalactic sources of the \textit{Planck} point
source catalogue (e.g., in the fields of H-ATLAS; \citealp{Eales2010}),
but confusion noise at the \textit{Planck} HFI channels induces uncertainties
in point source photometry that are too large for an accurate estimate
of the cross calibration of the two instruments, especially compared
to the level of precision that can be achieved with maps of diffuse
emission.

\begin{acknowledgements}
We thank Jean-Loup\,Puget for insightful discussions. BB is
particularly thankful to Marc-Antoine\,Miville-Desch\^enes for
providing us pre-release access to the Spider SPIRE data. BB
acknowledges the support of a CNES post-doctoral research grant.
\bfc{We thank the referee, Bernard Lazareff, for helpful comments that have led to improvements in the manuscript.}
The Planck Collaboration acknowledges the support of: ESA; CNES, and
CNRS/INSU-IN2P3-INP (France); ASI, CNR, and INAF (Italy); NASA and DoE
(USA); STFC and UKSA (UK); CSIC, MINECO, JA and RES (Spain); Tekes,
AoF, and CSC (Finland); DLR and MPG (Germany); CSA (Canada); DTU Space
(Denmark); SER/SSO (Switzerland); RCN (Norway); SFI (Ireland);
FCT/MCTES (Portugal); ERC and PRACE (EU). A description of the Planck
Collaboration and a list of its members, indicating which technical or
scientific activities they have been involved in, can be found at
\url{http://www.cosmos.esa.int/web/planck/planck-collaboration}.

\end{acknowledgements}

\bibliographystyle{aat} 

\bibliography{Planck_bib,astrobib}

\appendix

\section{Notes regarding absolute photometric calibration}
\label{app:calib}

The absolute calibration of the two instruments is described in Sect.\,\ref{sect:calib}.  Some supplementary details are provided here.

\subsection{\textit{Planck}/HFI}
\label{sect:hficalib}

HFI is a scanning instrument and its effective beams relevant to the
frequency maps are the convolution of (i) the optical response of the
telescope and feeds; (ii) the processing of the time-ordered data and
deconvolution of the time response; and (iii) the merging of several
surveys to produce frequency maps.  While accurate effective beams can
be recovered at each sky position, this is a time-consuming process.
However, the rms variations of the effective beam solid angle across the
sky are 0.79\,\% and 0.49\,\% at 545 and 857\GHz, respectively (see
Table 3 in \citealp{planck2014-a08}), which are negligible with
respect to the calibration uncertainty, and so ignoring the spatial
variation has a negligible effect for our study of diffuse emission
calibration.
Therefore, for convenience of implementation in our processing
pipeline we decided to use an effective Gaussian beam instead, as cited in Sect.\,\ref{sect:calib}.

Accurate knowledge of the spectral response is essential for
comparing surface brightness obtained through different filters or for
computing colour corrections.  Spectral responses for \textit{Planck}/HFI were
measured pre-flight for each detector and checked with ground and
in-flight data (see \citealp{planck2013-p03d}).  As noted in Sect.\,\ref{sect:calib}, these are known with sufficient accuracy not to impact the results in this paper.\\

The photometric calibration for the \textit{Planck}/HFI channels which are CMB dominated (100, 143, and
217~GHz) or have a strong enough CMB signal (353~GHz) is achieved through observations of the
dipole arising from the orbital motion of the \textit{Planck} spacecraft around the
Sun (the orbital dipole). The calibration accuracy is very high, 0.09\% and 0.07\% at
100 and 143 GHz, respectively \citep{planck2014-a09}. 

The 545~GHz channel and even more so the 857~GHz channel cannot be calibrated on
the orbital dipole, which is too weak with respect to Galactic dust
emission, and so, as described above, these were calibrated using Uranus
and Neptune. Nevertheless, given the high absolute calibration
accuracy at 100 and 143~GHz, the relative photometric
calibration of the 545~GHz channel with respect to those two channels
can be used to assess the accuracy of the 545~GHz planet-based absolute
calibration.

\citet{planck2014-a10} have used the dipole generated by the solar system motion with
respect to the CMB (the solar dipole) and the first two
acoustic peaks of the CMB fluctuations to measure the relative
calibration of all \textit{Planck} HFI channels, except 857~GHz, adopting the
average of 100 and 143~GHz as reference.
As summarized in Sect.\,\ref{sect:calib}, it was found that the planet-calibrated data at 545~GHz
has a relative calibration of $+2.3$\,\%$\,\pm\,$1.6\,\% using the solar
dipole and $+1.5$\,\%$\,\pm\,$1.8\,\% using the first two CMB peaks.
This is remarkable and suggests that the uncertainty of the planet-based absolute
calibration of the 545~GHz channel is not as great as the value of 6.1\,\%
cited in Sect.\,\ref{sect:calib}.  Recall that 5\,\% arose from the estimated absolute uncertainty of the planet model predictions,
\citep{planck2014-a09} which relates mostly to uncertainties on the
thermal profiles from \citet{Lindal1992}.
Even if the uncertainty were that large, the relative planet model uncertainty between these two HFI bands is expected to be of order 2\,\%.

The previous \textit{Planck} 2013 data release (PR1) also relied on the Neptune and
Uranus models for absolute calibration. Improvements in the data
processing and calibration procedure shifted the final absolute
calibration of frequency maps by $-1.8$\,\% and $-3.3$\,\% at 545 and
857\GHz, respectively (the 2015 PR2 maps are fainter).

\subsection{\Herschel/SPIRE}
\label{sect:spicalib}

The uncertainty of the conversion factor $K_{\text{PtoE}}$ (Sect.\,\ref{sect:calib}) is dominated by the 4\,\%
uncertainty of the effective beam solid angles, $\Omega_{\text{eff}}$.
We adopted values from \citet{Griffin:2013vn} ({\tt HIPE} Version 12).

Ongoing efforts to improve the modelling of the radial beam profile
have revised the values of $\Omega_{\text{eff}}$ to
$1804.3 \pm13\,\textrm{arcsec}^2$ ($\pm\,$0.7\,\%)
and
$831.4 \pm 3.8\,\textrm{arcsec}^2$ ($\pm\,$0.5\,\%)
for PLW and PMW, respectively. These amount to increases of just 
2.0\,\% and 1.1\,\%; furthermore, we note that
the stated uncertainties in $\Omega_{\text{eff}}$ \bfc{are now considerably} less than the 4\,\% cited above.  
Adopting the revised values of $\Omega_{\text{eff}}$ would decrease
the above conversion factors $K_{\text{PtoE}}$ and thus decrease the
brightness of the SPIRE maps and the SPIRE\,/\,HFI relative gains estimated
in this paper (Eqs.\,\ref{g545} and \ref{g857}). 

Although these values have been incorporated in {\tt HIPE} Version 14\footnote{\url{http://herschel.esac.esa.int/twiki/bin/view/Public/SpirePhotometerBeamProfile2}.}, nevertheless, until such time as revised beam solid angles are adopted
officially in the SPIRE Handbook, 
and to facilitate tracking of dependencies and changes, we felt it prudent to frame our results in terms of the cited values
from the version of {\tt HIPE} that we used (Version 12), consistent with the values
from the latest SPIRE Handbook, v2.5.  

\begin{figure}
\begin{center}
\includegraphics[width=8cm]{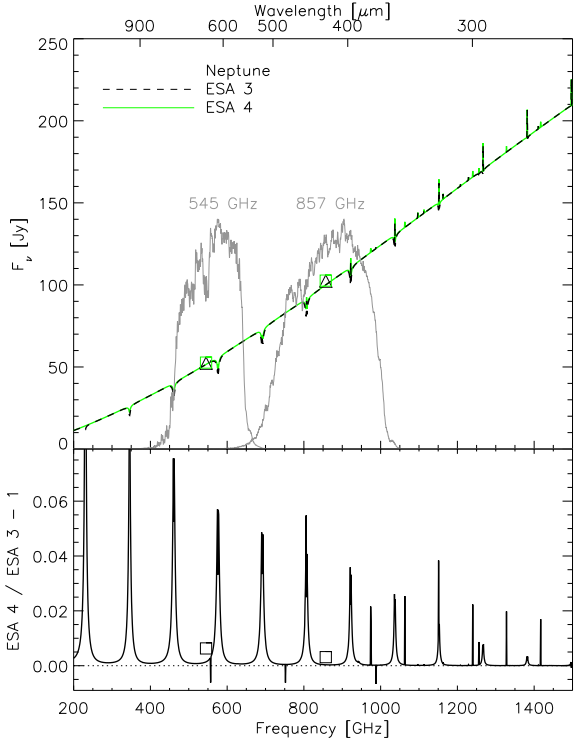}
\caption{{\it Top:} Comparison of the ESA\,3 (dashed black) and ESA\,4
(green) Neptune models. Corresponding flux densities in the 545 and
857\GHz\ HFI bands are shown as triangles (ESA\,3) and squares
(ESA\,4). {\it Bottom:} Relative difference of the two model spectra
and the flux densities.}
\label{fig:nepmodels}
\end{center}
\end{figure}

\subsection{Neptune models}
\label{sect:neptmod}

As seen in Fig.\,\ref{fig:nepmodels}, the main differences between the
ESA\,3 model of Nepture used for the HFI calibration, and the ESA\,4 model of Nepture used
for the SPIRE calibration rest in an updated treatment of the CO
absorption features and HCN emission lines. Several of these features
fall within the HFI and SPIRE bands and so affect estimates of
the cross-calibration factor as a systematic error. We computed HFI
equivalent Neptune flux densities for both models and find that, if
calibrated on ESA\,4 rather than ESA\,3, the HFI absolute calibration
factors on Neptune would increase by 0.63\,\% and 0.31\,\% at 545 and
857\GHz, respectively (see bottom panel of
Fig.\,\ref{fig:nepmodels}). Combined with the absolute photometric
calibration of HFI derived from Uranus model, these drop down to an increase
of 0.31\,\% and 0.16\,\% at 545 and 857\GHz, respectively.  We note that
an increase of the HFI calibration factors translates into an increase
of HFI brightness and thus a decrease of the SPIRE\,/\,HFI relative gains that
we estimate in this paper (Eqs.\,\ref{g545} and \ref{g857}).

\end{document}

%% file: Planck.tex
\def\setsymbol#1#2{\expandafter\def\csname #1\endcsname{#2}}
\def\getsymbol#1{\csname #1\endcsname}

\newbox\tablebox    \newdimen\tablewidth
\def\leaderfil{\leaders\hbox to 5pt{\hss.\hss}\hfil}
\def\endPlancktable{\tablewidth=\columnwidth 
    $$\hss\copy\tablebox\hss$$
    \vskip-\lastskip\vskip -2pt}

\def\tablenote#1 #2\par{\begingroup \parindent=0.8em
    \abovedisplayshortskip=0pt\belowdisplayshortskip=0pt
    \noindent
    $$\hss\vbox{\hsize\tablewidth \hangindent=\parindent \hangafter=1 \noindent
    \hbox to \parindent{$^#1$\hss}\strut#2\strut\par}\hss$$
    \endgroup}
\def\doubleline{\vskip 3pt\hrule \vskip 1.5pt \hrule \vskip 5pt}

\def\L2{\ifmmode L_2\else $L_2$\fi}

\def\DeltaT{\ifmmode \Delta T\else $\Delta T$\fi}
\def\deltat{\ifmmode \Delta t\else $\Delta t$\fi}
\def\fknee{\ifmmode f_{\rm knee}\else $f_{\rm knee}$\fi}
\def\Fmax{\ifmmode F_{\rm max}\else $F_{\rm max}$\fi}
\def\solar{\ifmmode{\rm M}_{\mathord\odot}\else${\rm M}_{\mathord\odot}$\fi}
\def\Msolar{\ifmmode{\rm M}_{\mathord\odot}\else${\rm M}_{\mathord\odot}$\fi}
\def\Lsolar{\ifmmode{\rm L}_{\mathord\odot}\else${\rm L}_{\mathord\odot}$\fi}
\def\inv{\ifmmode^{-1}\else$^{-1}$\fi}
\def\mo{\ifmmode^{-1}\else$^{-1}$\fi}
\def\sup#1{\ifmmode ^{\rm #1}\else $^{\rm #1}$\fi}
\def\expo#1{\ifmmode \times 10^{#1}\else $\times 10^{#1}$\fi}
\def\,{\thinspace}
\def\lsim{\mathrel{\raise .4ex\hbox{\rlap{$<$}\lower 1.2ex\hbox{$\sim$}}}}
\def\gsim{\mathrel{\raise .4ex\hbox{\rlap{$>$}\lower 1.2ex\hbox{$\sim$}}}}

\def\simprop{\mathrel{\raise .4ex\hbox{\rlap{$\propto$}\lower 1.2ex\hbox{$\sim$}}}}
\def\deg{\ifmmode^\circ\else$^\circ$\fi}
\def\pdeg{\ifmmode $\setbox0=\hbox{$^{\circ}$}\rlap{\hskip.11\wd0 .}$^{\circ}
          \else \setbox0=\hbox{$^{\circ}$}\rlap{\hskip.11\wd0 .}$^{\circ}$\fi}
\def\arcs{\ifmmode {^{\scriptstyle\prime\prime}}
          \else $^{\scriptstyle\prime\prime}$\fi}
\def\arcm{\ifmmode {^{\scriptstyle\prime}}
          \else $^{\scriptstyle\prime}$\fi}
\newdimen\sa  \newdimen\sb
\def\parcs{\sa=.07em \sb=.03em
     \ifmmode \hbox{\rlap{.}}^{\scriptstyle\prime\kern -\sb\prime}\hbox{\kern -\sa}
     \else \rlap{.}$^{\scriptstyle\prime\kern -\sb\prime}$\kern -\sa\fi}
\def\parcm{\sa=.08em \sb=.03em
     \ifmmode \hbox{\rlap{.}\kern\sa}^{\scriptstyle\prime}\hbox{\kern-\sb}
     \else \rlap{.}\kern\sa$^{\scriptstyle\prime}$\kern-\sb\fi}
\def\ra[#1 #2 #3.#4]{#1\sup{h}#2\sup{m}#3\sup{s}\llap.#4}
\def\dec[#1 #2 #3.#4]{#1\deg#2\arcm#3\arcs\llap.#4}
\def\deco[#1 #2 #3]{#1\deg#2\arcm#3\arcs}
\def\rra[#1 #2]{#1\sup{h}#2\sup{m}}

\def\dots{\relax\ifmmode \ldots\else $\ldots$\fi}
\def\WHzsr{\ifmmode $W\,Hz\mo\,sr\mo$\else W\,Hz\mo\,sr\mo\fi}
\def\mHz{\ifmmode $\,mHz$\else \,mHz\fi}
\def\GHz{\ifmmode $\,GHz$\else \,GHz\fi}
\def\mKs{\ifmmode $\,mK\,s$^{1/2}\else \,mK\,s$^{1/2}$\fi}
\def\muKs{\ifmmode \,\mu$K\,s$^{1/2}\else \,$\mu$K\,s$^{1/2}$\fi}
\def\muKRJs{\ifmmode \,\mu$K$_{\rm RJ}$\,s$^{1/2}\else \,$\mu$K$_{\rm RJ}$\,s$^{1/2}$\fi}
\def\muKHz{\ifmmode \,\mu$K\,Hz$^{-1/2}\else \,$\mu$K\,Hz$^{-1/2}$\fi}
\def\MJysr{\ifmmode \,$MJy\,sr\mo$\else \,MJy\,sr\mo\fi}
\def\MJysrmK{\ifmmode \,$MJy\,sr\mo$\,mK$_{\rm CMB}\mo\else \,MJy\,sr\mo\,mK$_{\rm CMB}\mo$\fi}
\def\microns{\ifmmode \,\mu$m$\else \,$\mu$m\fi}

\def\muK{\ifmmode \,\mu$K$\else \,$\mu$\hbox{K}\fi}
\def\microK{\ifmmode \,\mu$K$\else \,$\mu$\hbox{K}\fi}
\def\muW{\ifmmode \,\mu$W$\else \,$\mu$\hbox{W}\fi}
\def\kms{\ifmmode $\,km\,s$^{-1}\else \,km\,s$^{-1}$\fi}
\def\kmsMpc{\ifmmode $\,\kms\,Mpc\mo$\else \,\kms\,Mpc\mo\fi}
\providecommand{\sorthelp}[1]{}